\documentclass[namedreferences]{SolarPhysics}
\usepackage[optionalrh,solaenum]{spr-sola-addons} % For Solar Physics
\usepackage{graphicx} % For eps figures, newer & more powerfull
\usepackage{sidecap}
\usepackage{color} % For color text: \color command
\usepackage{url} % For breaking URLs easily trough lines
\newcommand{\FeXXI}{Fe {\sc xxi}}

\newcommand{\FeXIX}{Fe {\sc xix}}
\newcommand{\FeXX}{Fe {\sc xx}}
\newcommand{\todash}{\,--\,}
\newcommand{\kms}{km~s$^{-1}$}

\begin{document}

\begin{article}

\begin{opening}

\title{Multiwavelength Observations of an Eruptive Flare: Evidence for Blast Waves and
Break-out}
\author{Pankaj~\surname{Kumar}$^{1,2}$\sep
        D. E.~ \surname{Innes}$^{2}$
                  }
\runningauthor{P. Kumar et al.}
\runningtitle{Evidence for blast waves and break-out}
\institute{$^{1}$ Korea Astronomy and Space Science Institute (KASI),
Daejeon,
      305-348, Republic of Korea. email: \url{pankaj@kasi.re.kr}}
\institute{$^{2}$ Max-Planck Institut f\"{u}r Sonnensystemforschung,
37191 Katlenburg-Lindau, Germany. email: \url{innes@mps.mpg.de}}
%%%%%%%%%%%%%%%%%%%%%%%%%%%%%%%%%%%%%%%%%%%%%%%%%%%%%%%%%%%%%%%%%%%%%%%%
\begin{abstract}
Images of an east-limb flare on 3 November 2010 taken in the 131~\AA\ channel
of the {\it Atmospheric Imaging Assembly} onboard the {\it Solar Dynamics
Observatory} provide a convincing example of a long current sheet below an erupting plasmoid,
 as predicted by the standard magnetic reconnection
model of eruptive flares. However, the 171~\AA\ and 193~\AA\ channel images
hint at an alternative scenario. These images reveal that large-scale waves
with velocity greater than 1000~\kms\ propagated alongside and ahead of the
erupting plasmoid.
 Just south of the plasmoid, the waves coincided with
type-II radio emission, and to the north, where the waves propagated along
plume-like structures, there was increased decimetric emission. Initially the
cavity around the hot plasmoid expanded. Later, when the erupting plasmoid
reached the height of an overlying arcade system, the plasmoid structure
changed, and the lower parts of the cavity collapsed inwards. Hot loops
appeared alongside and below the erupting plasmoid. We consider a scenario in
which the fast waves and the type-II emission were a consequence of a flare
blast wave, and the cavity collapse and the hot loops resulted from the break-out
of the flux rope through an overlying coronal arcade.

\end{abstract}
\end{opening}
%-------------------------------------------------

%%%%%%%%%%%%%%%%%%%%%%%%%%%%%%%%%%%%%%%%%%%%%%%%%%%%%%%%%%%%%%%%%%%%%%%%%%%%%%%%%%%%%%%%%%%%%
\section{Introduction}

The flare and flux rope eruption observed on the east limb of the Sun on 3
November 2010 showed many of the expected features of eruptive flare models
\cite{Car64,Sturrock66,Hira85,KP76}. In particular, a fast moving plasmoid,
interpreted as a flux rope, appeared above the flare site with what looked
like a long, hot current sheet below
\cite{Reeves11,Cheng11,Savage12,Hannah12}. Hot, inflowing plasma was observed
below the plasmoid seemingly causing it to accelerate and drive a shock,
indicated by type-II radio emission, in the low corona
\cite{Bain12,Zimovets12}. The eruption also triggered transverse oscillations
in the surrounding loop systems \cite{White12a,White12b}.

To understand the connection between the loops and the erupting plasmoid, we
took a careful look at the relation between the hot (131~\AA) and cooler (193
and 171~\AA) plasma emission. This revealed fast waves along plumes outside
the hot cavity that were also headed by enhanced radio emission, suggesting
that the flare caused shock-like signatures considerably beyond the plasmoid,
possibly due to a flare blast wave. Although blast waves at the onset of
energetic flares are predicted (\opencite{Vrsnak00b}, \citeyear{Vrsnak00a}; \opencite{vrsnak2008}; \opencite{magdalenic2008}, \citeyear{magdalenic2010}), they
have proved difficult to observe. One major problem was that they propagate
with speeds greater than 1000~\kms, and therefore required high image cadence
and a large field of view to be detected in images. Thus it was not until
images from the {\it Atmospheric Imaging Assembly} (AIA; \opencite{Lemen12})
became available that there was a reasonable chance of seeing them. Blast
waves have however been invoked to explain the high (greater than 500~\kms)
Doppler shifts observed in \FeXIX\ and \FeXX\ ultraviolet emission lines seen
across a significant fraction of flaring active region coronae
\cite{Ietal01,Tothova11}, and also in events exhibiting type-II radio bursts
where there was only slow or no associated coronal mass ejection (CME)
\cite{Vrsnal06,Magdalenic12}.

The 3 November 2010 eruption is also significant because of its well-observed
inflows below the erupting plasmoid which have been interpreted as inflows to
a current sheet. As pointed out by \inlinecite{Hannah12}, these inflows are
hot (8\todash14~MK), not cool, as predicted by the standard model. If the
fast waves result from a flare blast wave, then the plasmoid acceleration may
have occurred before the inflows, and this leads us to consider the
possibility that the inflows are a consequence of flux rope break-out through
the overlying coronal arcade \cite{Antiochos99,Karpen12}. Break-out leads to
reconnection in the corona, the outward acceleration of the flux rope, and
the formation of postflare loops on the side and below the break-out site, as
well as inflows to a current sheet below the flux rope. We therefore take a
further look at the structure of the flare heated plasma seen in the 131 and
94~\AA\ images to understand the dynamics beyond the regions of the
previously identified inflows and current sheet.

In this paper we first show the evidence for fast propagating waves and their
relation to enhanced radio emission. We then compare the timing of break-out
with the onset of plasma inflow below the plasmoid, and the structures seen
in the 131 and 94~\AA\ channel emissions. A summary sketch of the observed
structures and our interpretation is given in the final section.

%%%%%%%%%%%%%%%%%%%%%%%%%%%%%%%%%%%%%%%%%%%%%%%%%%%%%%%%%%%%%%%%
\section{Observations and Results}
The eruption occurred in active region NOAA 11121 which appeared on the
eastern limb (N38E18) on 3 November 2010. The C4.9 flare
 was first seen in STEREO-B {\it Extreme UltraViolet Imager} (EUVI; \opencite{Howard08})
  images at 12:10~UT and had a hard X-ray
 onset at 12:13~UT and peak at 12:14~UT \cite{Zimovets12}.
The event, observed by the AIA onboard the {\it Solar Dynamics Observatory}
(SDO; \opencite{Pesnell2012}), the {\it Reuven Ramaty High Energy Solar
Spectroscopic Imager} (RHESSI; \opencite{lin2002}), the {\it Nan\c{c}ay
Radioheliograph}  (NRH; \opencite{Kerdraon1997}), and the San Vito Radio Solar
Telescope has caught attention because several of its features resemble the
standard flare model and has presented a plausible case for a piston-driven
shock \cite{Bain12,Zimovets12}.

In this analysis, we discuss AIA 131, 94, 193, and 171~\AA\ image sequences
taken with a cadence of 12~s. Most of the images are shown after the
subtraction of a pre-flare image taken at approximately 12:10~UT ({\it i.e.}
base-difference images) to reveal the changes caused by the flare. To find
the location of the radio emission, we have used high-cadence radio images of
the Sun from the NRH, which operates in the frequency range of 160\todash 435
MHz and probes the coronal plasma at heights $\le$1.5 {\it R$_\odot$}
\cite{Kerdraon1997}.

\begin{figure}
\includegraphics[width=\textwidth]{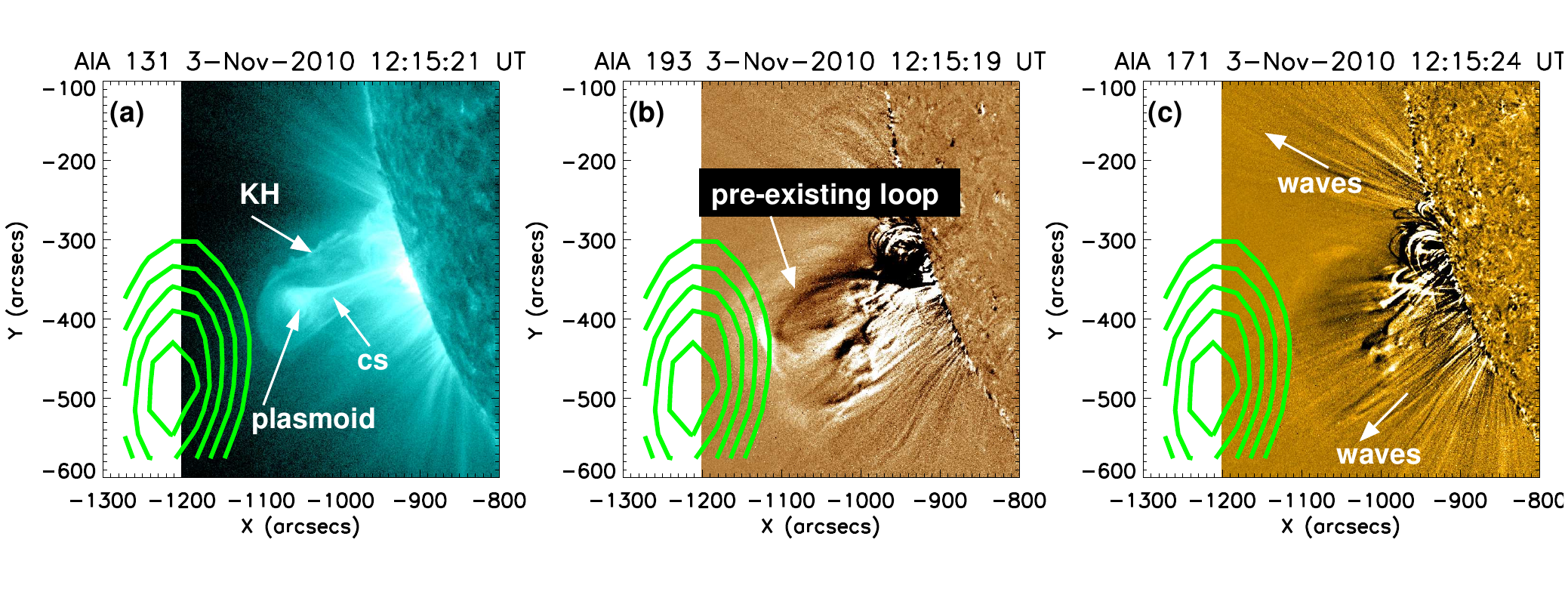}
\caption{SDO/AIA images of the 3 November 2010 eruption with radio 270~MHz contours
at 50\%, 60\%, 70\%, 80\%, 90\%\ of maximum. (a) 131~\AA\ intensity with previously
identified current sheet (CS), plasmoid, and Kelvin-Helmholtz roll-ups (KH) marked.
(b) 193~\AA\ base difference showing the position of the pre-existing loop as a
dark loop. (c) 171~\AA\ base difference. The 171~\AA\ and 131~\AA\ base images were taken at
12:09:00 and 12:09:09 UT, respectively.}
\label{over_3nov}
\end{figure}

An overview of several of the features discussed in previous papers is shown
in Figure~\ref{over_3nov}. The 131~\AA\ image, dominated by \FeXXI\ emission,
shows a hot, $\approx10$~MK, plasmoid directly above and connected to the
flare site by a long, thin thread, which has been interpreted as a current
sheet (CS). The plasmoid swept up the surrounding plasma into a bright rim of
cooler, $\approx2$~MK, 193~\AA\ emitting plasma \cite{Reeves11,Cheng11}.
Radio observations revealed a type-II radio burst near the leading edge of
the expanding plasmoid system \cite{Bain12,Zimovets12}. Behind the plasmoid,
converging towards CS, were hot plasma inflows \cite{Savage12}. The ripples
along the northern flank of the hot cavity have been interpreted as
Kelvin-Helmholtz (KH) roll-ups \cite{Foullon11}. Important features for our
later discussion are also illustrated in the 193~\AA\
(Figure~\ref{over_3nov}b) and 171~\AA\ (Figure~\ref{over_3nov}c)
base-difference images. Here dark represents pre-flare active region
structures that have disappeared due to heating or eruption. For example, the
dark loop in the 193~\AA\ image corresponds to a pre-flare coronal loop or
arcade. The 171~\AA\ base-difference image shows that there was enhanced
emission both south and north of the hot, plasmoid cavity.

%%%%%%%%%%%%%%%%%%%%%%%%%%%%%%%%%%%%%%%%%%%%

%%%%%%%%%%%%%%%%%%%%%%%%%%%%%%%%%%%%%%%%%%%%
\subsection{Early EUV Waves}

\begin{SCfigure}
\centering
\includegraphics[width=8cm]{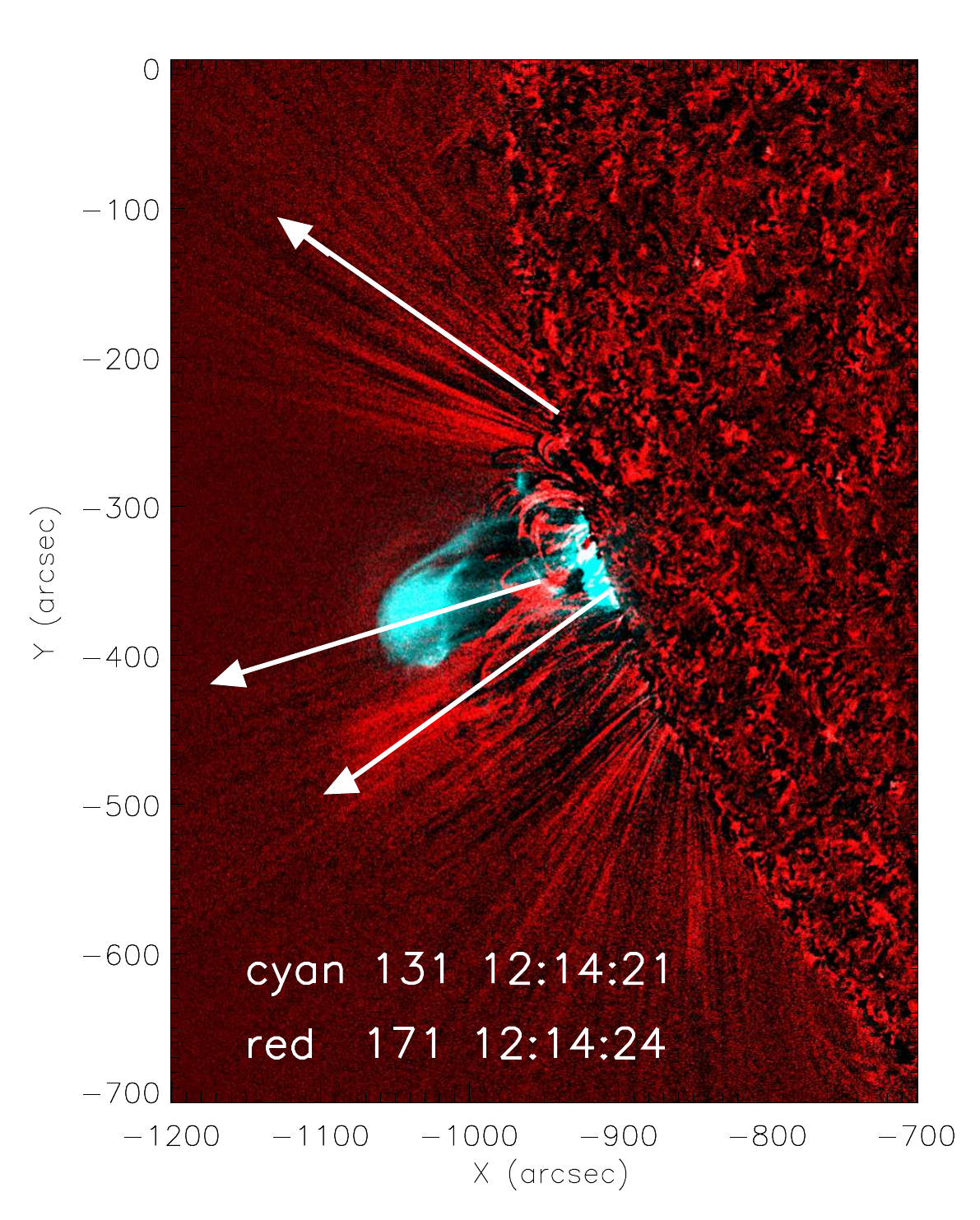}
\caption{Composite of 131~\AA\ (cyan) and 171~\AA\ (red) base difference
images showing a broad enhanced front of 171~\AA\ emission just south of the
hot plasmoid, observed at 131~\AA. The white lines show the position of the
space-time images in Figure~\ref{st_waves}. This is one frame from the movie
{\sl 131\_171.avi}} \label{waveimages}
\end{SCfigure}

\begin{figure}
\includegraphics[width=\linewidth]{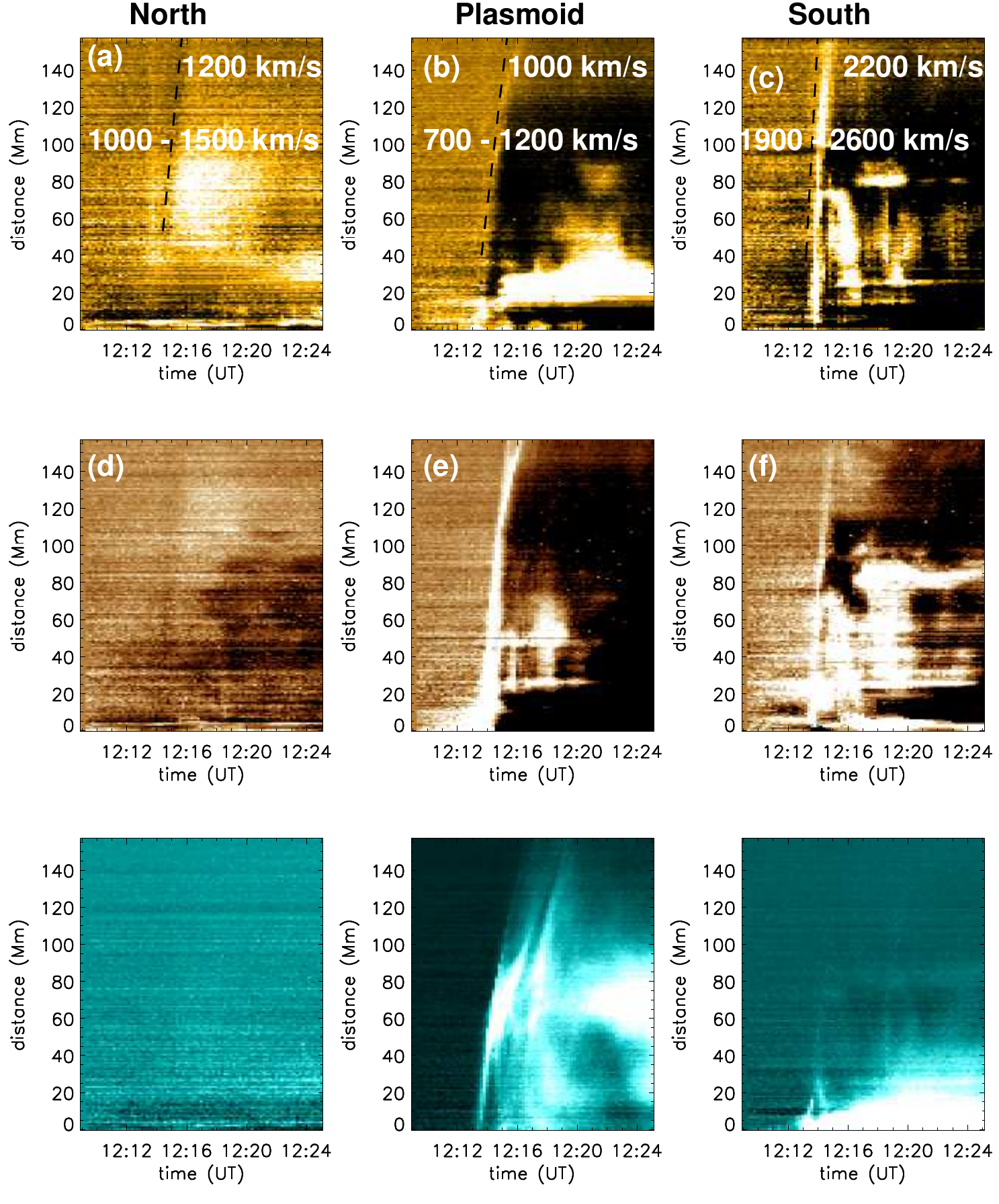}
\caption{Space-time base difference images of 171\AA\ (top row), 193\AA\ (middle) and
131\AA\ (bottom) emission, along the three white lines shown in Figure~\ref{waveimages}.
The left column is along the northern line, the middle along the plasmoid and the
right column along the southern line. Dashed white lines show the track for a front with the
labeled speed.}
\label{st_waves}
\end{figure}

During the early stage of the flare/flux-rope eruption, fast waves were
observed in sequences of 171~\AA\ images. Figure~\ref{waveimages} is a
composite of AIA 171~\AA\ and 131~\AA\ base-difference images showing the
regions and directions of the fast waves. This is one frame from the movie
linked to the figure in the online version. In Figure~\ref{waveimages}, the
EUV wave, south to the plasmoid is visible as enhanced 171~\AA\ emission extending
beyond the height of the 131~\AA\ plasmoid. The enhanced emission seen in the
north at this time is probably due to resonance scattering of extreme
ultraviolet (EUV) flare emission because the whole region brightened
simultaneously and propagating waves were not seen here until 12:15:00~UT.
The wave speeds can be deduced from the space-time images shown in
Figure~\ref{st_waves}. This method computes the time when there is an abrupt
change in the intensity at each spatial position. Then a linear function is
fitted to the values to obtain the velocity. When the front changes speed, it is
sensitive to the start and end distance used to fit the wave front. For
example, the front in Figure~\ref{st_waves}b changed from 1200~\kms\ at the
start to 700~\kms\ at the end. On the image, we draw the line corresponding to
the best estimate of the velocity and provide the velocity range. The waves to
the south had a very high plane-of-sky speed, about 2200~\kms. The speed was
about twice that of the plasmoid and the waves along the plumes to the north.
Both the southern waves and the plasmoid eruption started at 12:13~UT, a
couple of minutes before the waves to the north. The composite AIA 131 and
171~\AA\ base difference movie clearly shows the bright EUV wavefront south
of the 131~\AA\ plasmoid, and the propagation of the EUV wave along the
plumes, north of the flare site.

%%%%%%%%%%%%%%%%%%%%%%%%%%%%%%%%%%%%
%%%%%%%%%%%%%%%%%%%%%%%%%%%%%%%%%%%%%%

%%%%%%%%%%%%%%%%%%%%%%%%%%%%%%%%%%%%%%%%%%%%%%%%%%%%%%%%%%%%%%%%%%%%%%%%%%%%%%%%%%%%%%%%%%%
\subsection{Radio Emission}
 The radio emission has been discussed in
the context of the RHESSI hard X-ray emission by \inlinecite{Zimovets12} and
\inlinecite{Bain12}. At flare onset there was a short burst of hard X-ray
(25\,--\,50~keV) and microwave (3000\,--\,5000~MHz) emission, peaking at
$\approx$12:14~UT. A few minutes later, there was a decimetric type-II radio
burst that showed distinct band-splitting, indicating two outward-moving
sources. \inlinecite{Zimovets12} interpreted these as the up- and downstream
regions of a shock but it is also possible that the splitting is due to waves
along two different density structures. \inlinecite{Zimovets12} found that one source
started slightly off-set (100~Mm) south of the hot plasmoid trajectory, and
had a speed 2240~\kms, whereas the other was in line with the plasmoid and
had a velocity 1500~\kms.

\begin{figure}
\centerline{
\includegraphics[width=0.5\textwidth]{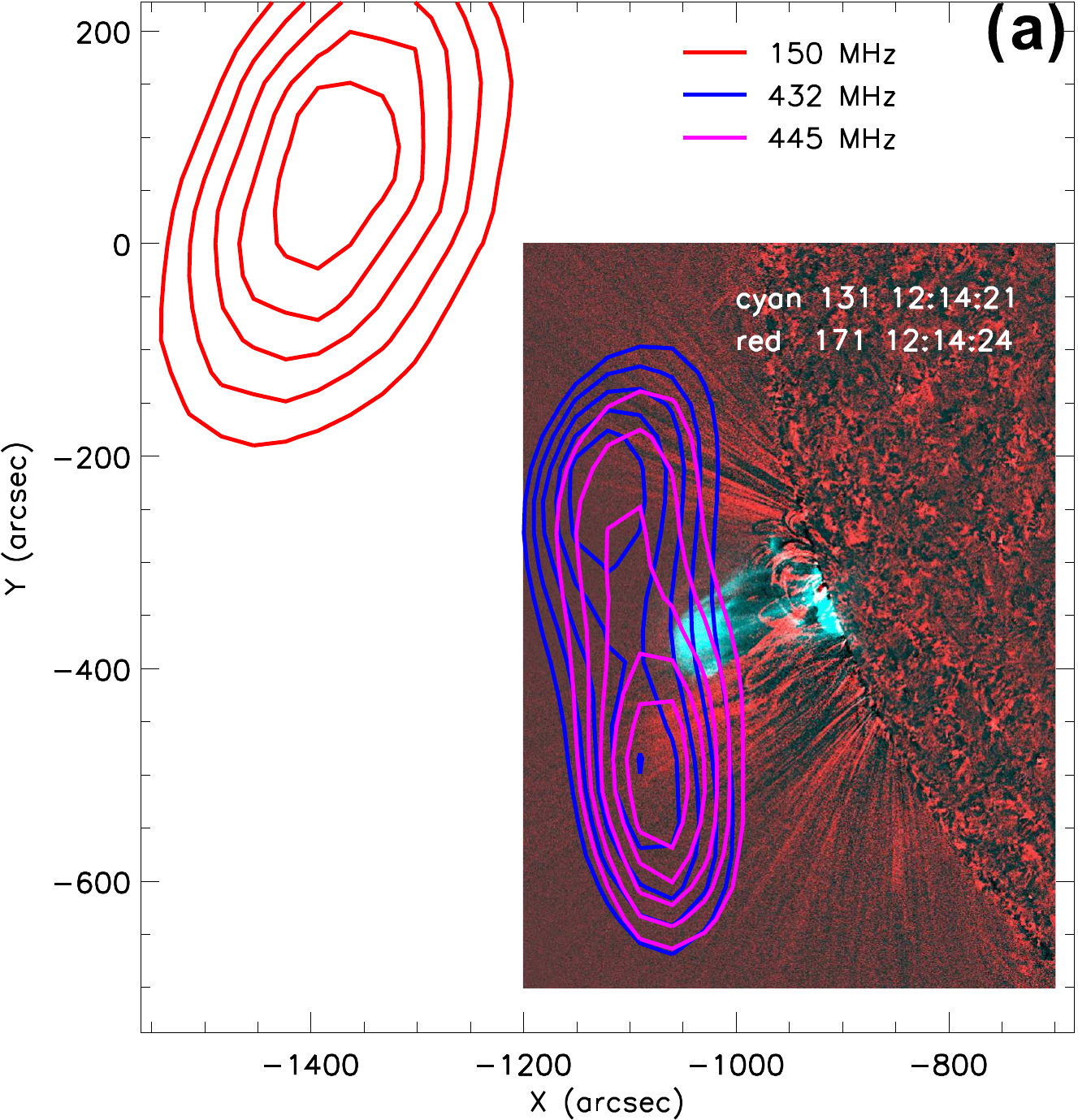}
\includegraphics[width=0.5\textwidth]{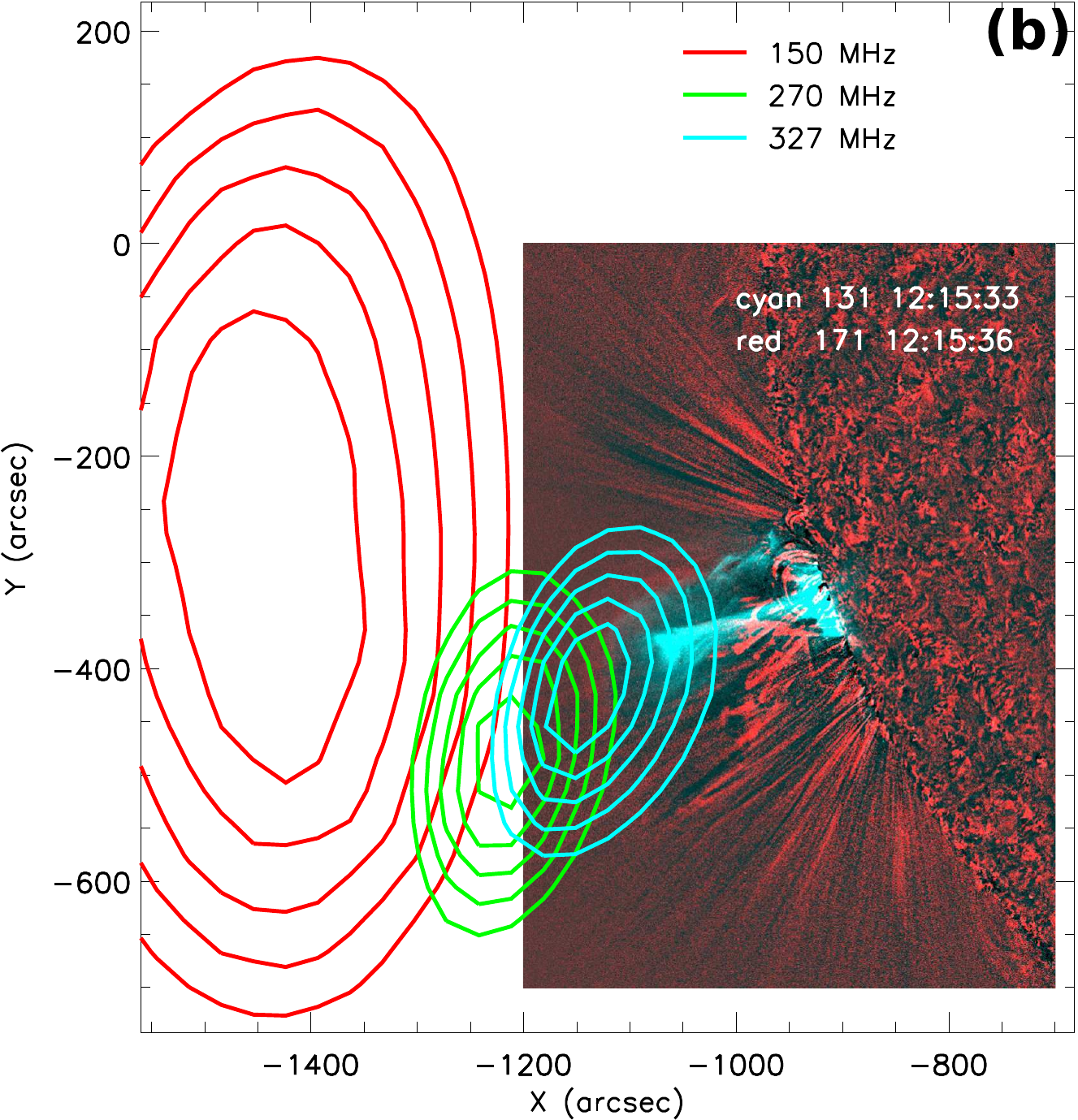}
}
\centerline{
\includegraphics[width=0.5\textwidth]{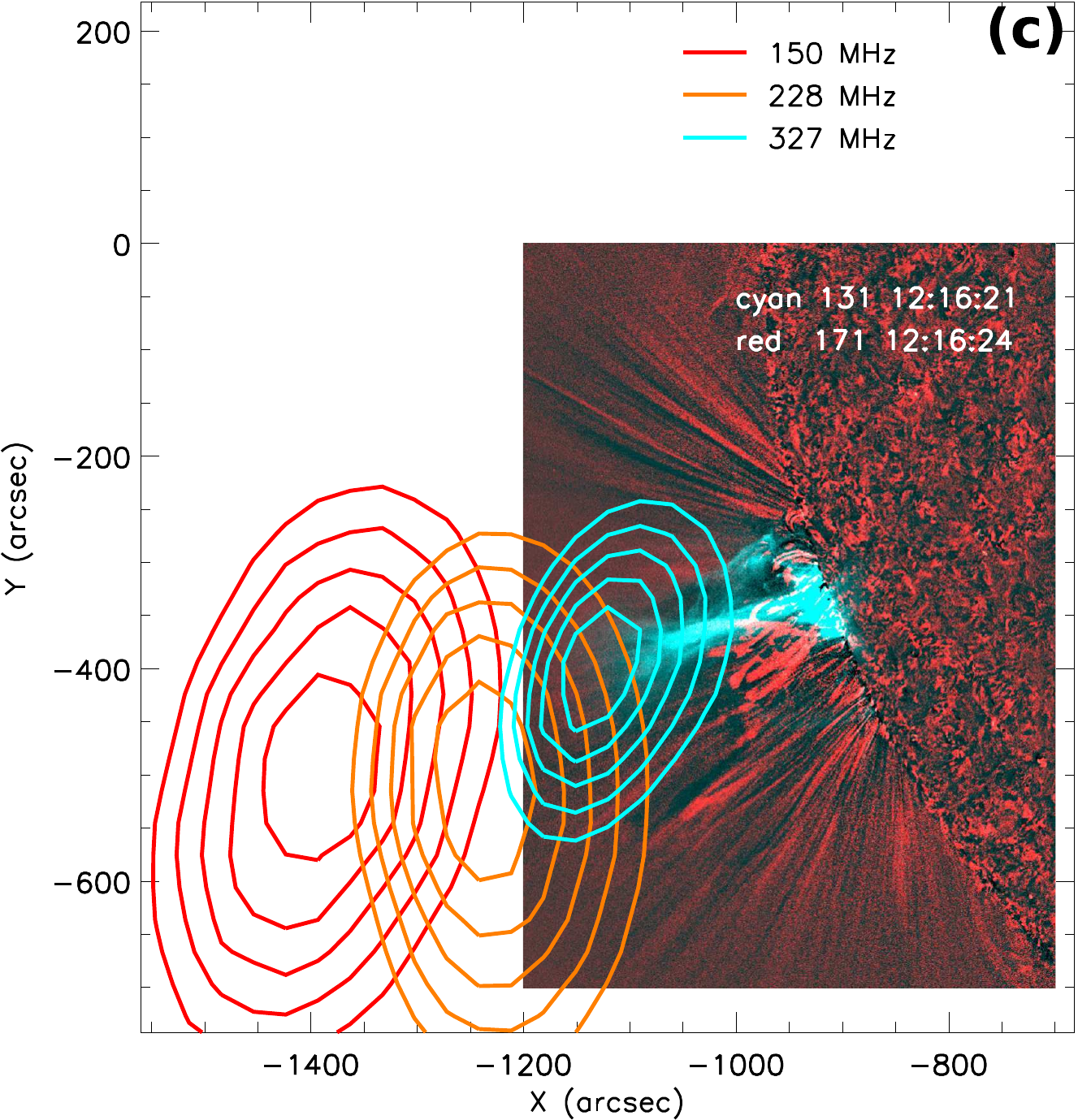}
\includegraphics[width=0.5\textwidth]{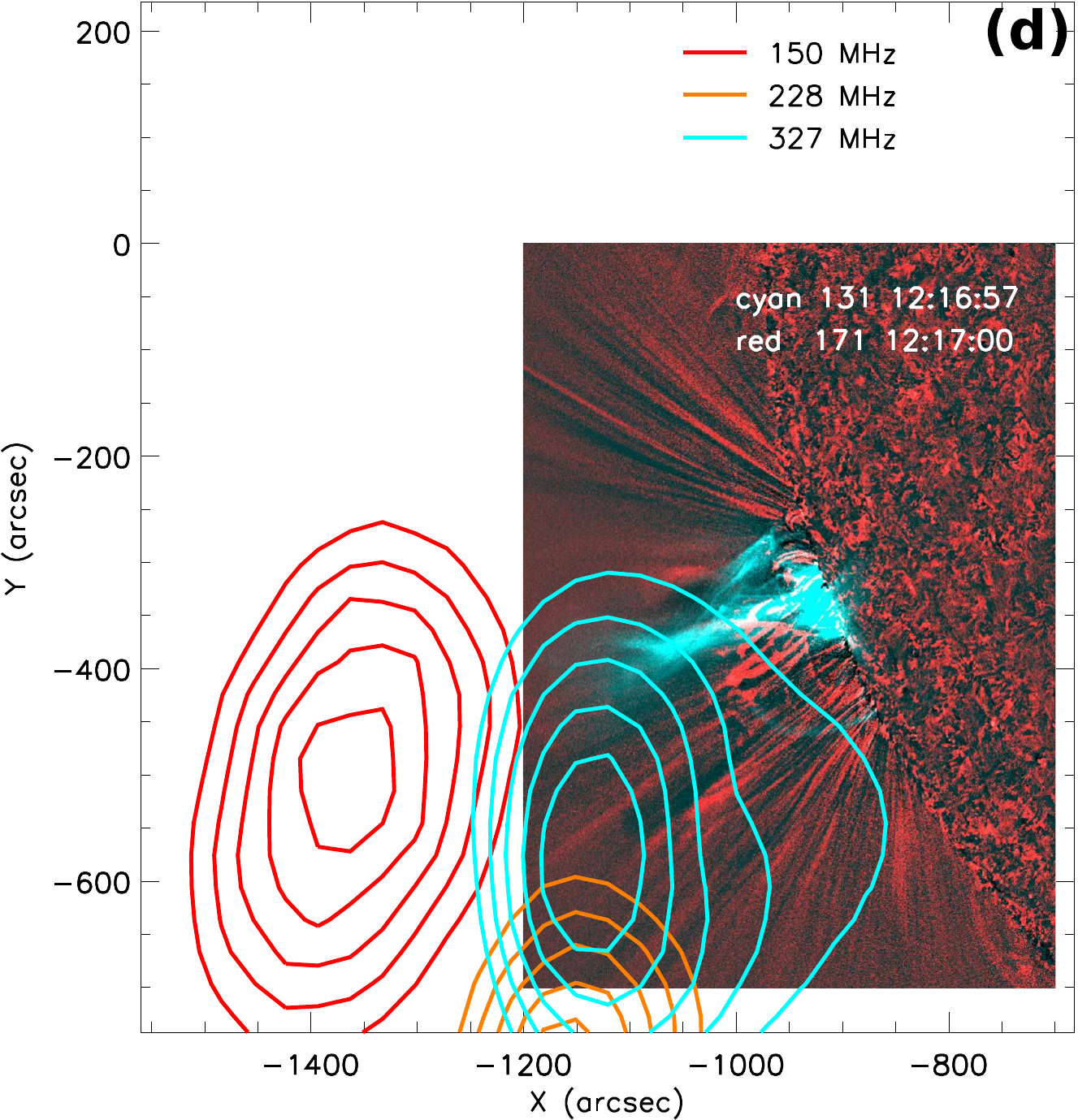}
}
\centerline{
\includegraphics[width=0.52\textwidth]{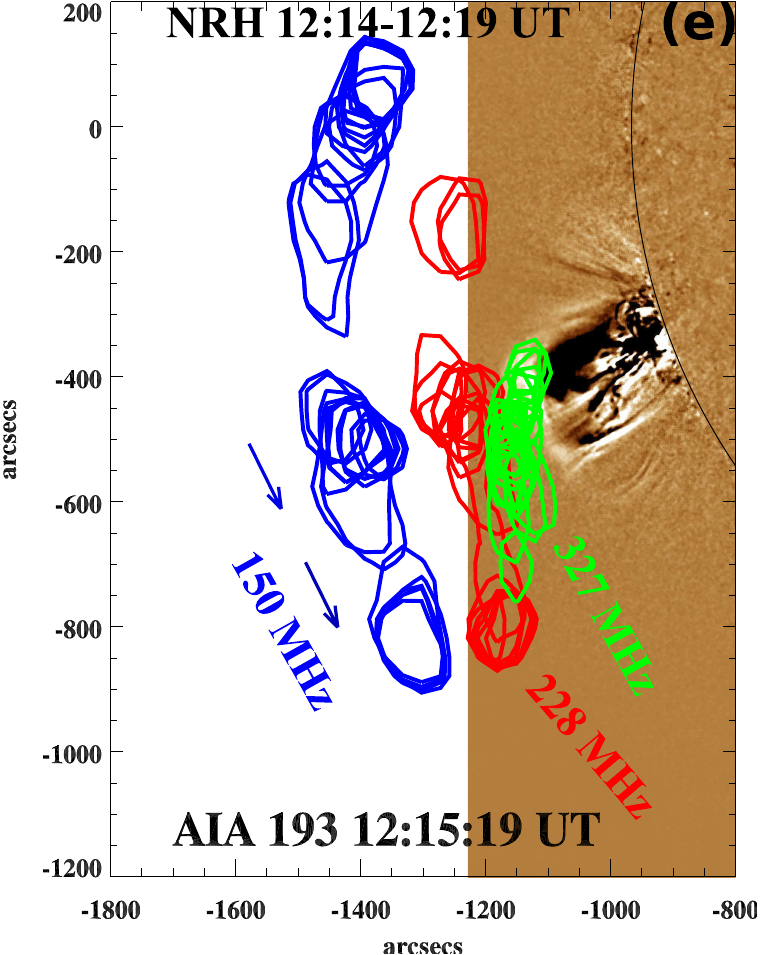}
\includegraphics[width=0.49\textwidth]{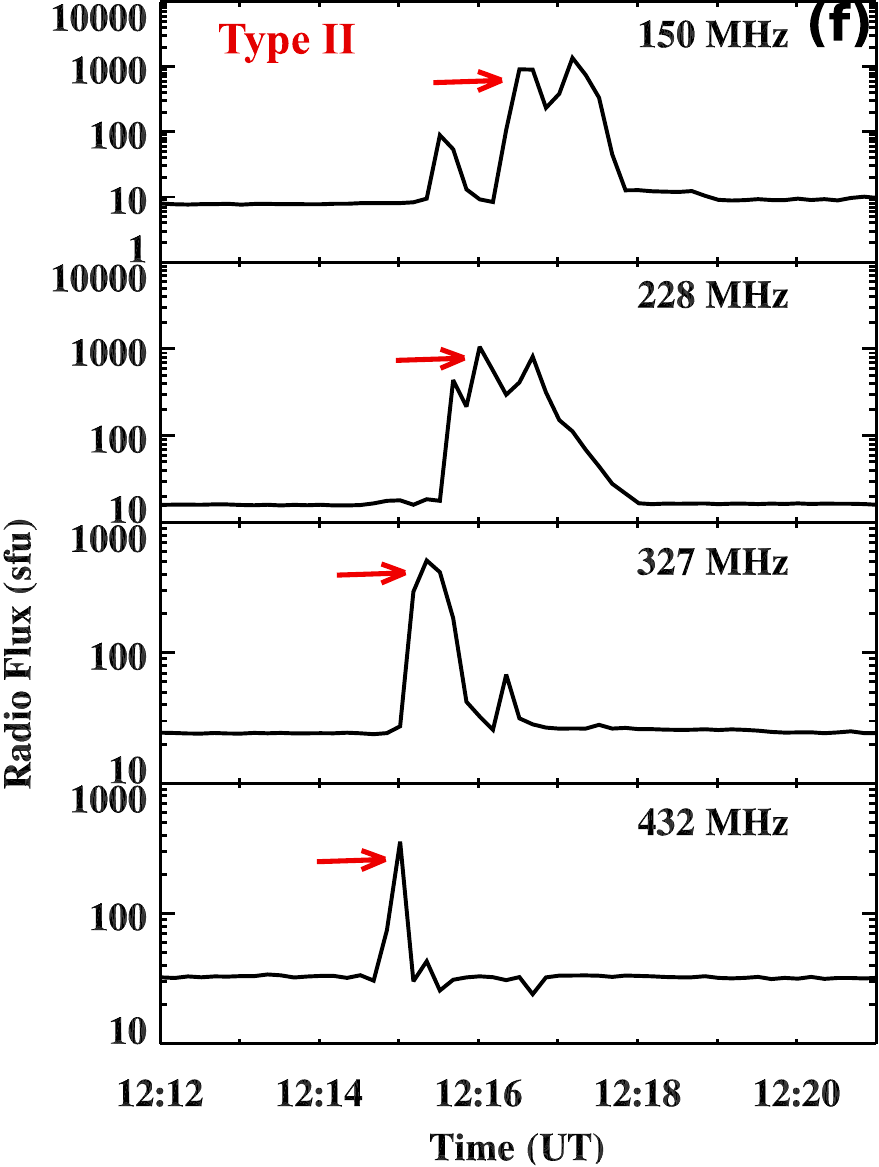}
}
\caption{Evolution of radio emission in the context of the EUV waves
and hot plasmoid. The bottom panels show the drifting type II sources indicated by arrows.}
\label{radio}
\end{figure}

In Figure~\ref{radio}, we compare the timing and sites of the radio emission
with the waves. Figure~\ref{radio}a, shows that early on there was enhanced
445 and 432~MHz emission across a large portion of the active region, with
the most intense emission on either side of the erupting plasmoid. The
southern radio source is the low frequency component tracked by
\inlinecite{Zimovets12} and estimated to have a velocity 2240~\kms. As can be
seen in the images, it coincided with the trajectory of the fast, 2200~\kms\
waves inferred from the 171~\AA\ images, so it is possible that the waves
excited this component of the radio emission.

The second component which was labelled the high frequency component by
\inlinecite{Zimovets12}, was seen along the trajectory of the hot plasmoid.
It did not appear until 1~min later at 327~MHz (Figure~\ref{radio}b), and had
a speed of 1500~\kms, which was faster than the hot plasmoid plane-of-sky
speed, and close to the speed of the 211~\AA\ leading edge \cite{Zimovets12}.
Subsequently, emission in front of the plasmoid dominated at frequencies at
and below 327~MHz, and both high and low frequency radio components appeared
over an extended region above the plasmoid. The radio emission may then have
been produced by up- and downstream regions of a shock or by waves along
different structures as suggested by earlier images.

In Figure~\ref{radio}, we also show 150~MHz contours. This emission comes
from waves in lower density plasma than the higher frequency emission, so it
is seen later and higher in the corona when caused by an outward propagating
source. On this day, there was a 150~MHz frequency source at the head of the
northern plumes (Figure~\ref{radio}a) where the northern 171~\AA\ fast waves
were observed. This source had been there long before the flare and it
brightened at the time of the flare (Figure~\ref{radio}f). The NRH maps at
the time of the early 150~MHz source brightening are shown in
Figure~\ref{radio}b. The center of the 150~MHz emission moved in front of the
hot plasmoid at 12:16:24~UT, as part of the type-II burst. Later, as the
flare evolved the centroid of the low frequency (150\todash 327~MHz) emission
moved south (Figures~\ref{radio}d and \ref{radio}e). The location of the radio sources with
respect to the eruption site is shown in Figure~\ref{radio}e, where we have
overlaid radio source (150, 228, and 327 MHz) contours on the AIA 193~\AA\
running difference image of the eruption at 12:15:19 UT. This image shows
that from 12:14-12:19 UT, the radio sources drifted southward (indicated by an
arrow). Figure 4f displays the integrated radio flux
profiles (in sfu) at four frequencies obtained from NRH. It is interesting to see
the drifting type II source (indicated by red arrows) at all frequencies.

%%%%%%%%%%%%%%%%%%%%%%%%%%%%%%%%%%%%%%%%%%%%%%%%%%%%%
%%%%%%%%%%%%%%%%%%%%%%%%%%%%%%%%%

\subsection{Break-out}
Break-out occurred when the flux-rope broke through the overlying cavity
system. Several structural changes appeared at the time of beak-out. The
overlying loop system disrupted and the plasmoid structure changed. At the
same time, a hot thread of emission appeared below and attached to the
plasmoid, and the earlier expansion of the cavity changed to contraction.

\begin{figure}
%\centerline{
%\includegraphics[width=0.33\textwidth]{dtosc_171.pdf}
%\includegraphics[width=0.33\textwidth]{dtosc_193.pdf}
%\includegraphics[width=0.33\textwidth]{dtosc_131.pdf}
%}
\includegraphics[width=\textwidth]{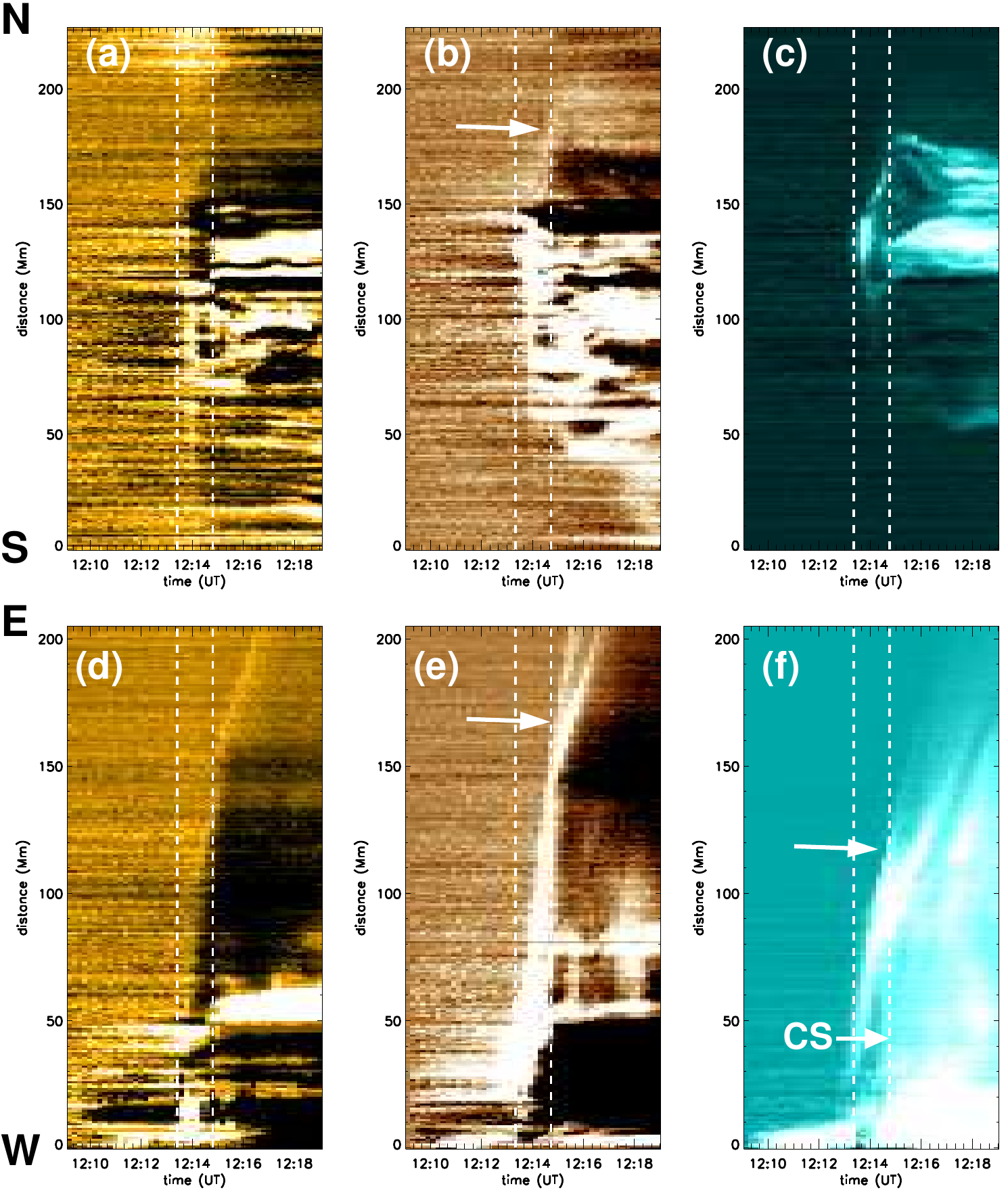}

 \caption{Time-distance images along lines shown in Figure~\ref{breakout_image}b.
 (a) 171~\AA\ along the S-N line; (b) 193~\AA\ along the S-N line; (c) 131~\AA\ along
 the S-N line; (d) 171~\AA\ along the W-E line; (e) 193~\AA\ along the W-E line;
 (f) 131~\AA\ along the W-E line. The first dashed line indicates flare onset
 and the second break-out time. The arrows point to fronts mentioned in the text.}
 \label{breakout_time}
\end{figure}

\begin{figure}
\centerline{
\includegraphics[width=0.5\textwidth]{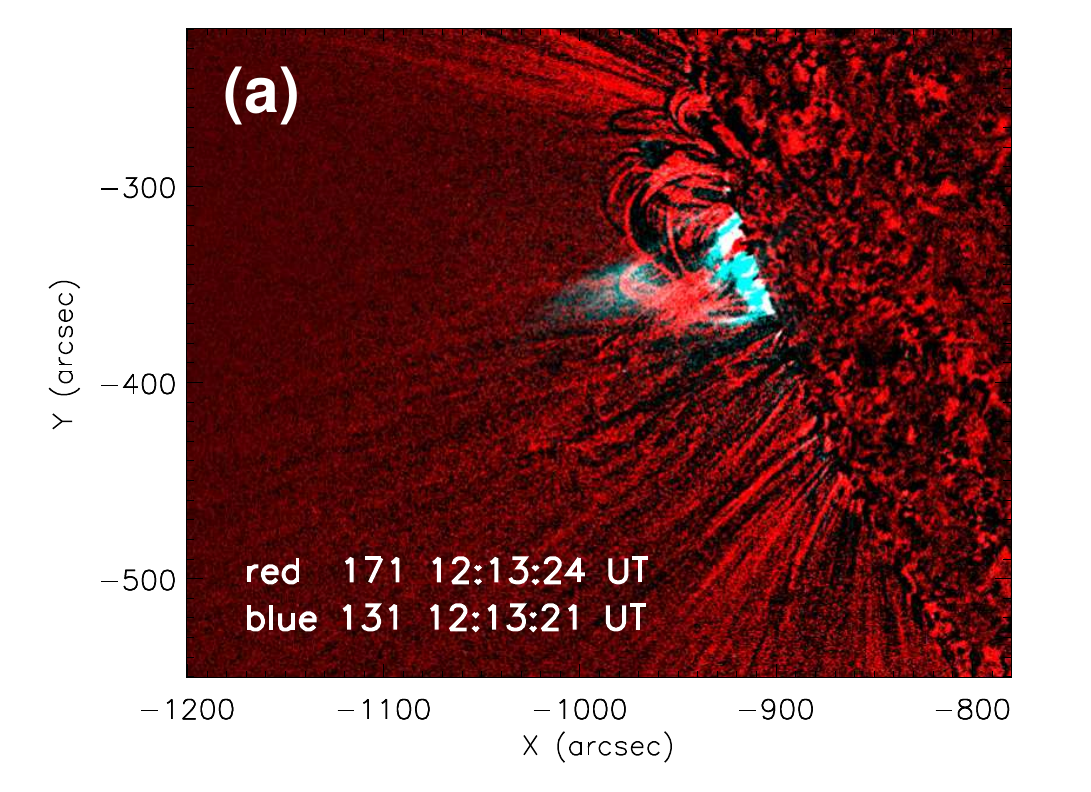}
\includegraphics[width=0.5\textwidth]{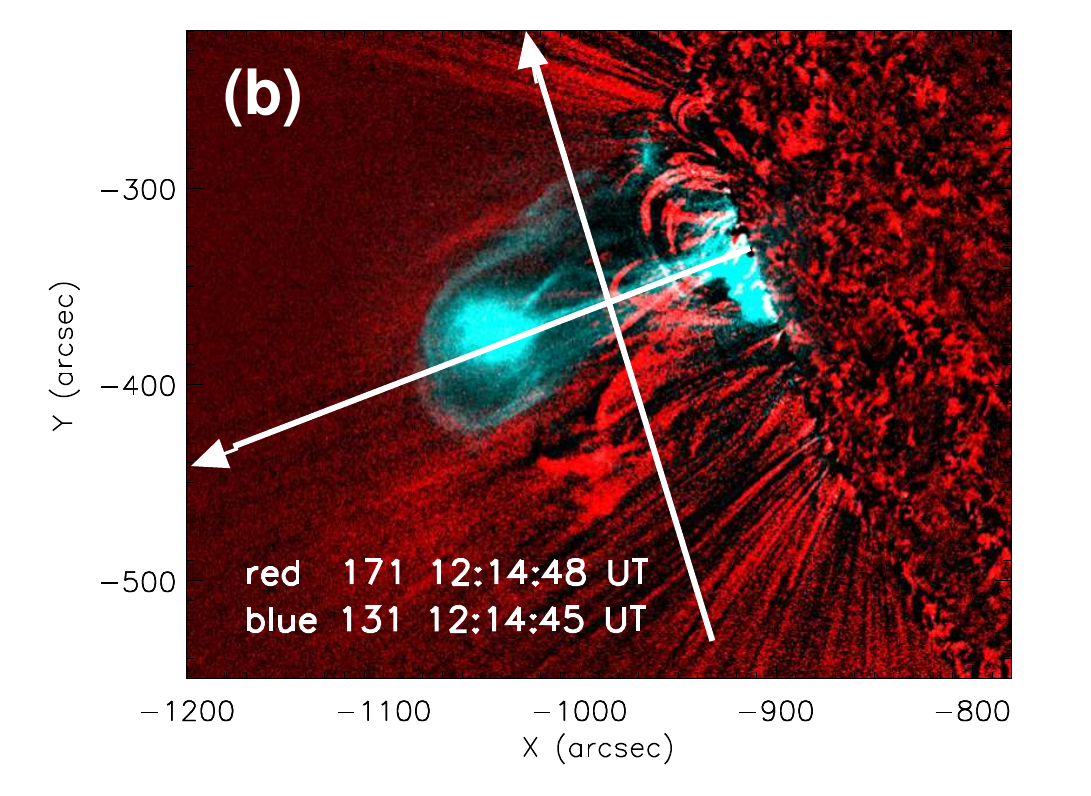}
}
\centerline{
\includegraphics[width=0.5\textwidth]{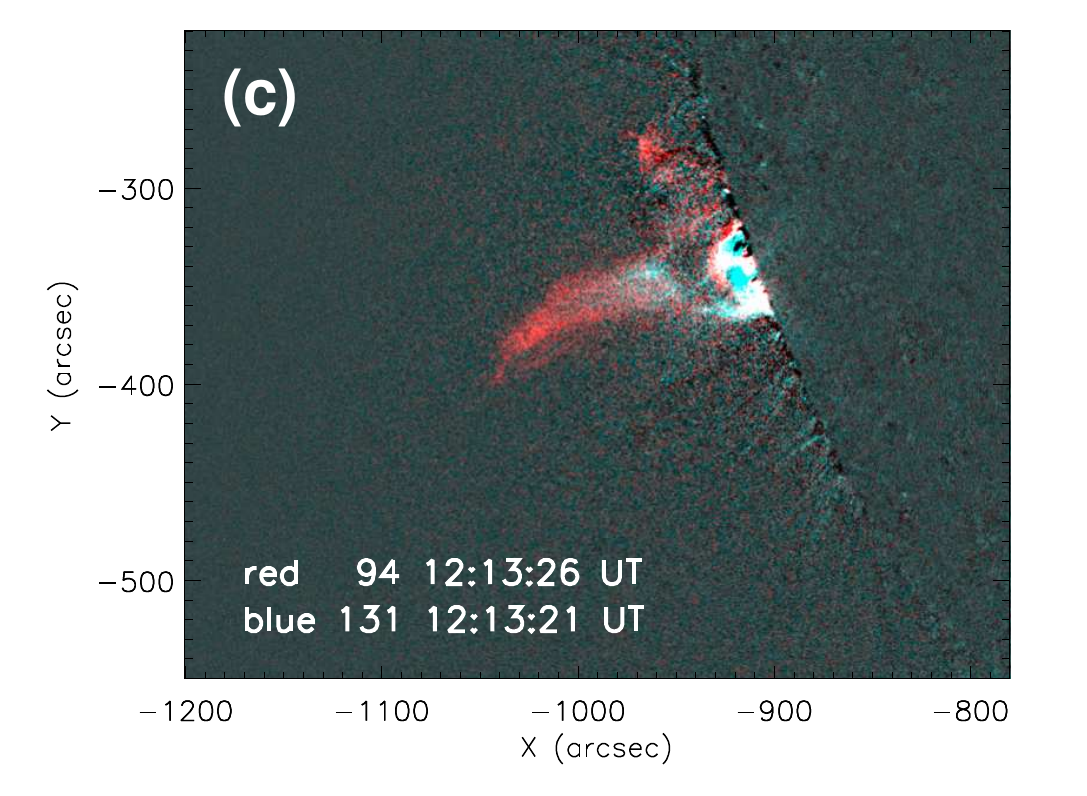}
\includegraphics[width=0.5\textwidth]{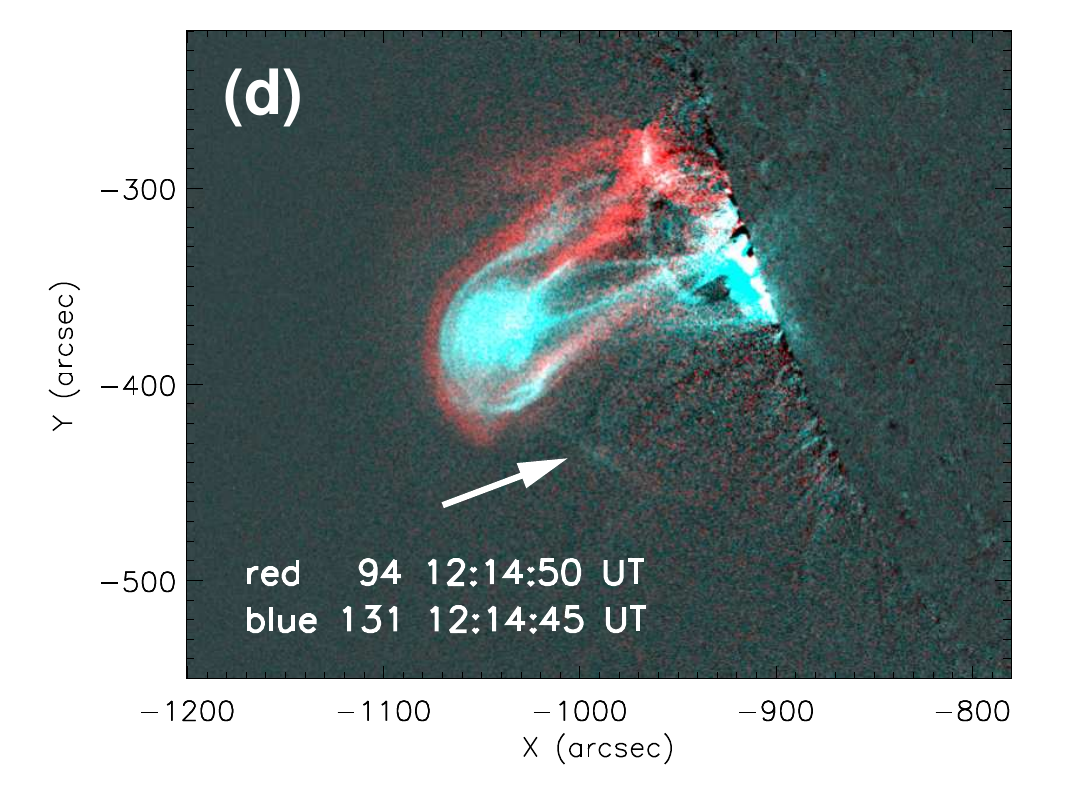}
}
\centerline{
\includegraphics[width=0.5\textwidth]{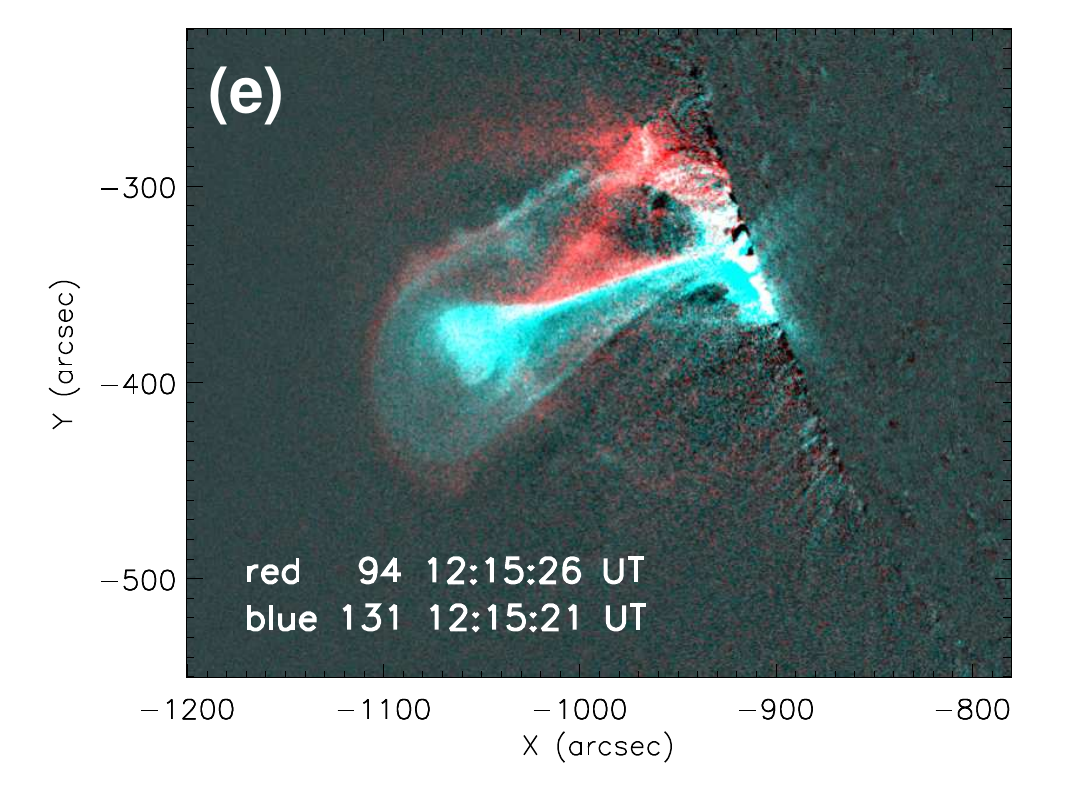}
\includegraphics[width=0.5\textwidth]{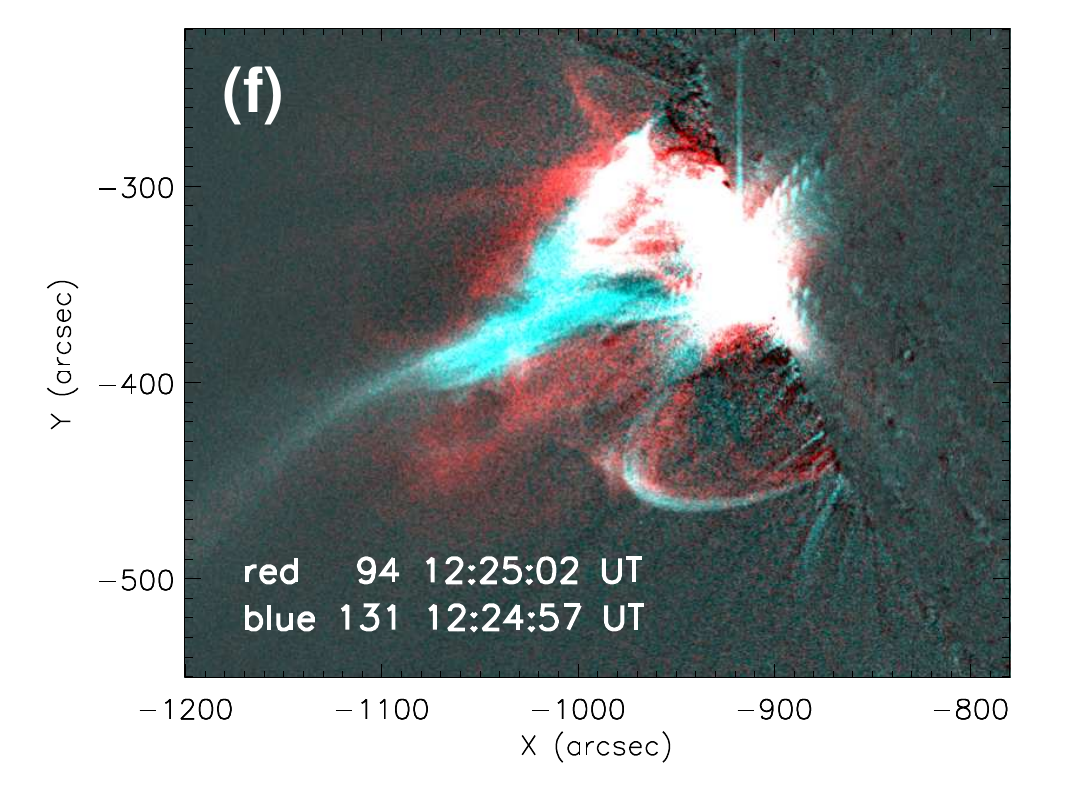}
}
\caption{Evolution of hot plasma in the context of oscillating loops seen in
171~\AA\ channel. (a and b) 171~\AA\ and 131~\AA\ composites; (c - f) 94~\AA\ and 131~\AA\
 composites.
On panel (b) the white lines show the position of the time-distance images in
Figure~\ref{breakout_time}. The top row shows frames from the movie {\sl 131\_171.avi} and
the other four frames are from the movie {\sl 131\_94.avi}. }
\label{breakout_image}
\end{figure}

Some of these features are illustrated in Figure~\ref{breakout_time}. These
are time-distance images taken across (top) and along (bottom) the direction
of propagation of the erupting flux rope, as shown in
Figure~\ref{breakout_image}b. These two perpendicular directions have been
chosen because they allow comparison of the expanding and contracting cavity
below the plasmoid and the outward propagating waves. The first dashed line
marks the flare onset time, and the second dashed line the time of break-out,
as deduced from the change in structure of the plasmoid leading edge, shown
in the bottom row of Figure~\ref{breakout_time}.

The images at 171 and 193~\AA\ show the broad bright front of the EUV waves
generated at onset, followed by dimming north and south of the flux rope
(Figure~\ref{breakout_time}a). Loop brightening and oscillations appeared
immediately behind the front over most the central region. The early wave
fronts are not visible in images of the hotter, 131~\AA\ emission. The
131~\AA\ image on the top row shows the expansion and contraction of the flux
rope cavity, especially in the north, where the cavity edge lies along the
inner edge of the 193~\AA\ faint front indicated with an arrow in
Figure~\ref{breakout_time}b. At the time of the second dashed line the cavity
expansion changes to contraction.
%%%%%%%%%%%%%%%%%%%%%%%%%%%%%%%%%%%%%%%%%%%%%%%%%%%%%
%%%%%%%%%%%%%%%%%%%%%%%%%%%%%%%%%%%%%%%%%%%%%%%%%%%%%%
%%%%%%%%%%%%%%%%%%%%%%%%%%%%%%%%%%%%%%%%%%%%%%%%%%%%%%%%%%%%%%%%%%%%%%%%%%%%%%%%%%%%%%%%%%%%%%%%%%%%%%%%%%%%%

As mentioned above, the second dashed line in Figure~\ref{breakout_time} was
placed to coincide with the time that the plasmoid changed shape. The leading
front in the 193 and 131~\AA\ space-time images on the bottom row split at
the time of the second dashed line: the plasmoid slowed down while the fast front
seen in 193~\AA\ images continued through the corona. At this time, the
193~\AA\ loop opened up and the bright 131~\AA\ flux rope changed from
circular to concave \cite{Cheng11,Zimovets12}, and it looks as though there
was a fundamental change in the plasmoid propagation. We associate this
sudden change with break-out. We also note that according to
Figure~\ref{breakout_time}f, the current sheet formed about 30~s (two frames)
before the second dashed line. Thus the current sheet formed while the cavity
was still expanding.

The structural changes are best seen in the accompanying movies which are two
colour composites of the 131 and 94~\AA, and the 131 and 171~\AA\ emissions.
A few frames from the movies are shown in Figure~\ref{breakout_image}. The
top two rows show the emission at onset and the last row shows the hot
post-eruption loops as seen in the 131/94~\AA\ images. The selected images
show that initially (around 12:13:24~UT), a bright tongue of 94~\AA\ emission
appeared above the 131~\AA\ flare emission and 171~\AA\ waves. The tongue
then separated and was filled with the plasmoid or flux rope of hotter
131~\AA\ emission. In the image shown in Figure~\ref{breakout_image}d, the
131~\AA\ plasmoid appears to be encased in a cocoon of 94 and 131~\AA\
emission. On the northern edge of the cocoon, close to the limb bright
94~\AA\ emission appeared. This seems to have been early, low-lying loops in
the bright flare arcade seen in Figure~\ref{breakout_image}f. On the south of
the plasmoid cavity, a region of faint, hot emission is just visible. The
131~\AA\ image at 12:14:45~UT was also shown in \inlinecite{White12b}, where
they comment on the faint line of hot emission, indicated by an arrow,
running from the plasmoid to the site of the hot loop seen later in 131~\AA\
images, and shown in Figure~\ref{breakout_image}f. The 94~\AA\ emission,
especially the emission seen at 12:25:02~UT, suggests that this line of
131~\AA\ emission outlined the top of a hot-loop arcade that formed at the
time of break-out.

%%%%%%%%%%%%%%%%%%%%%%%%%%%%%%%%%%%%%%%%%%

\subsection{STEREO View of the Flare Site}
\begin{figure}
\centerline{
\includegraphics[width=\textwidth]{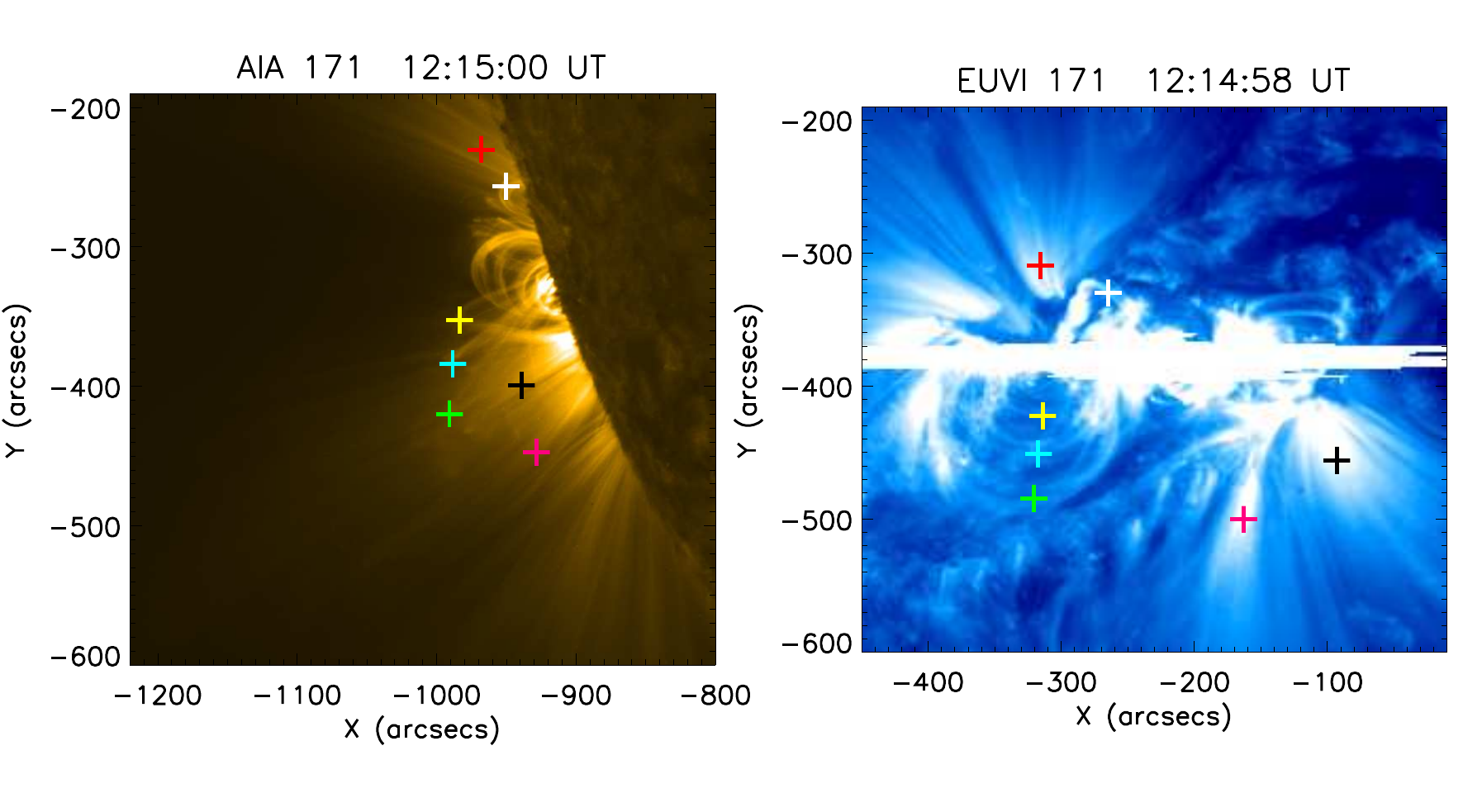}
}
\caption{Co-temporal images of AIA 171 and STEREO-B 171~\AA\ for identification
of loops and plumes. Crosses of the same colour identify the same points
in the two images.}
\label{aia_euvi}
\end{figure}
This flare that was seen over the east limb from Earth, was observed near
disk center in the STEREO-B/EUVI (hereafter EUVI-B) images. \inlinecite{Savage12} have
identified the main flare footpoints and flaring loops in the EUVI-B 195~\AA\
images. At 195~\AA, the cool plumes and loops are less visible than at
171~\AA. Therefore, in Figure~\ref{aia_euvi}, we show the 171~\AA\ EUVI-B and
AIA images. Unfortunately, the EUVI-B image is partially saturated by the
flare emission, so that loops that were below and along the line-of-site to
the erupting flux rope can not be seen. However the loops that were to the
north and south, and the sites of the plumes are clearly visible.

To find the positions of structures from the different perspectives, we have
used the SolarSoft routine scc$\_$measure.pro \cite{thompson2006}. We
selected a point on the AIA image and then identified the same feature on the
epipolar line drawn on the EUVI-B image. The same positions are marked with
the same color cross in the two images. The EUVI image shows that the plumes
were more than 100~arcsecs from the flare site. We note that there was an
arcade of large loops to the south and a faint (in 171~\AA) arcade of loops
to the north linking the flare footpoints, as identified by
\inlinecite{Savage12}.

%%%%%%%%%%%%%%%%%%%%%%%%%%%%%%%%%%%%%%%%%%%%%%%%%%%%%%%%%%%%%%%%%%%%%%%%%%%%%%
%%---------------------------------------------------------------------------

%%%%%%%%%%%%%%%%%%%%%%%%%%%%%%%%%%%%%%%%%%%%%%%%%%%%%%%%%%%%%%%%%%%%%%%%%%%%%%%%%%%%%%%%%%
\section{Summary and Discussion}
\begin{figure}
\centerline{
\includegraphics[trim = 1cm 5cm 1cm 2cm, clip, width=\textwidth]{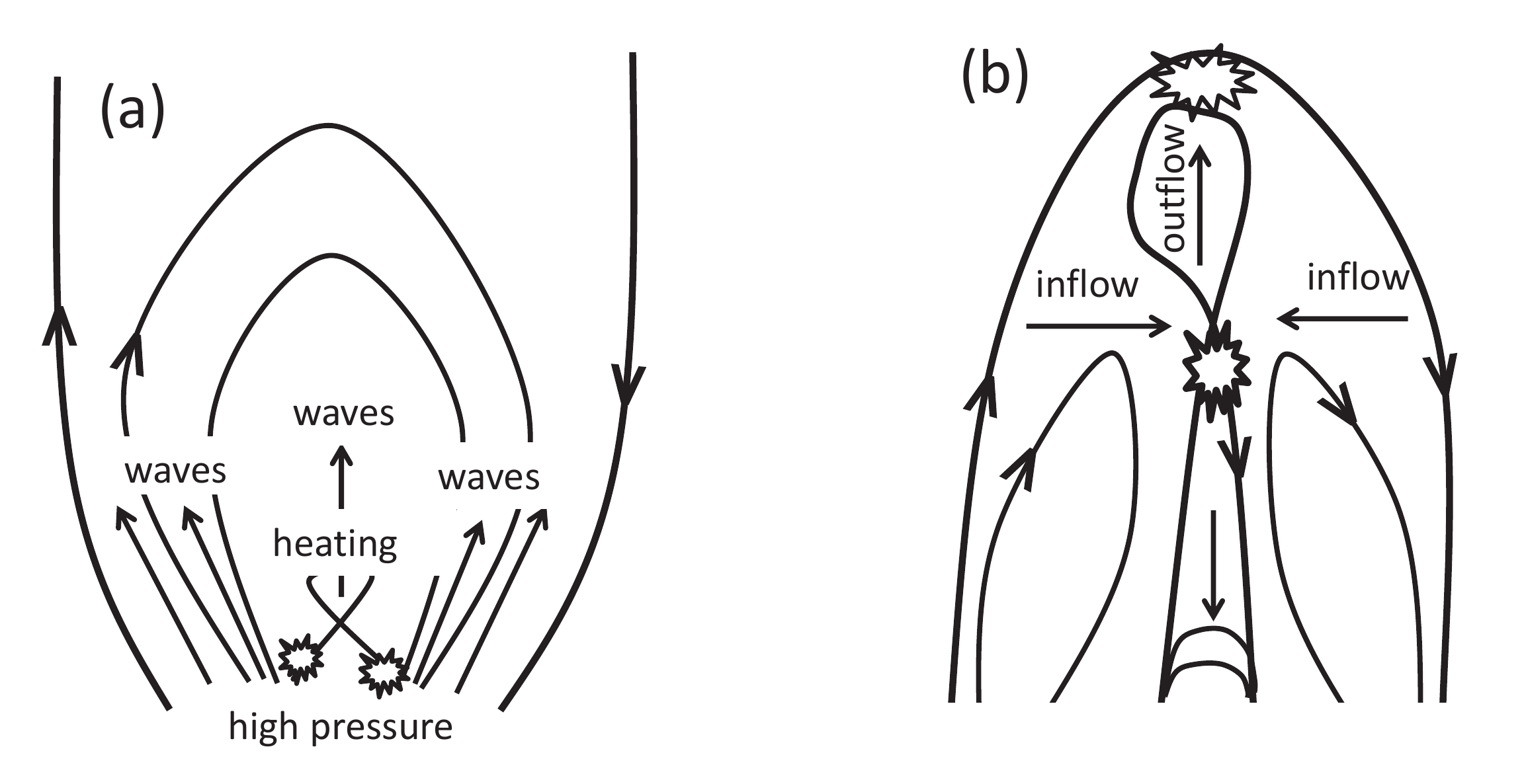}
}
\caption{Sketch of loop system at the time of the (a) flare onset and (b) break-out.
Reconnection site are shown as explosions.}
\label{sketch}
\end{figure}

We have taken a second look at the waves and structures observed by SDO/AIA
during a flare eruption that occurred on 3 November 2010, on the east limb of
the Sun as seen from Earth. At flare onset, fast 1000\todash2000~\kms\ EUV
waves were seen propagating along cool (1~MK) plume-like structures that were
situated about 100~arcsec from the flare site, and headed by decimetric
emission. These waves were traveling ahead and alongside a fast, flare-heated
plasmoid seen in 131~\AA\ images. Loops ahead of the plasmoid, heated to
about 5~MK, were seen in the 94~\AA\ channel images. Figure~\ref{sketch}a, is
a simplified sketch of the various features.

When the hot plasmoid reached the apex of the pre-flare loops, the leading
edge of the plasmoid accelerated away from the main core. At the same time,
hot loops were seen below and on the side of the plasmoid and the region
below collapsed inward to fill the space evacuated by the outward moving
plasmoid, as sketched in Figure~\ref{sketch}b.

Initially flaring in the low corona/chromosphere caused a rapid pressure
increase, flux rope destabilization, cavity expansion, and high energy
particles. Waves propagated outward in all directions. Heating was confined
to the inner flux-rope structure and its cavity. The rapidly rising flux rope
soon encountered the overlying arcade. Reconnection with the overlying field
led to break-out \cite{Antiochos99,Karpen12}. The formation of high-lying,
hot loops alongside and just below the plasmoid and the plasma inflows were a
consequence of break-out. Post-flare loops appeared below the plasmoid, close
to the limb. Since the eruption occurred over the limb as seen from Earth, it
is not possible to determine the exact flare site, the magnetic field
configuration, or the flare trigger. However, the STEREO 171~\AA\ image
(right panel of Figure \ref{aia_euvi}) shows the complex quadrapolar magnetic
field configuration associated with a possible null point above the flare
site, which are needed for the breakout reconnection.

This scenario is very different from previous ones for this event in two
respects. First, we discovered fast waves at the onset of the flare and this
led to the interpretation that a blast wave rather than a piston-driven shock
caused the type-II radio emission. Second, we noted that the inflows occurred
when the plasmoid structure changed and the over-lying 193~\AA\ loop
structure opened up and thus conclude that the inflows are as a consequence
of break-out. In this scenario, the bright 131~\AA\ thread that was below the
plasmoid could be either the current sheet or a newly-formed loop seen edge
on. The observation of hard X-ray emission near the plasmoid \cite{Bain12}
favour the current sheet interpretation but the width of the emission and its
stability suggest that it may be a newly formed loop similar to those seen on
either side of the plasmoid.

We could not observe the formation of a piston-driven shock in the AIA 211
and 193~\AA\ running difference images as observed in other flux-rope
eruption events \cite{ma2011,kozarev2011,gopalswamy2012}. The flux-rope is
heavily decelerated (mean speed $\approx$241~\kms) in the interplanetary (IP)
medium, and there was no piston-driven shock in the IP space (e.g., IP
type-II radio burst). Moreover, there is no strong lateral expansion of the flux rope,
that could be sufficient to drive the shock at the flanks of the CME
\cite{patsourakos2012}. We observed fast waves simultaneously at both sides
of the flux-rope (along the plumes), therefore it is also possible that the
pressure pulse over-ran the flux rope causing acceleration of the plasmoid in the low corona.
Flare-heated plasma was observed at the flare site, in the flux-rope and
along loops outlining the surrounding cavity, suggesting significant heating
throughout the flux rope cavity, not just along the central current sheet
(see movie {\sl 131\_94.avi}).

In conclusion, we reported a unique multiwavelength observation of the fast
propagating EUV waves in AIA 171~\AA\ and associated radio emissions ahead of
these waves, which supports the flare blast-wave scenario as a possible
candidate to generate high frequency type-II radio bursts. However, further
analysis of such events will be helpful to understand the eruption processes
associated with the generation of these fast waves.

%%%%%%%%%%%%%%%%%%%%%%%%%%%%%%%%%%%%%%%%%%%%%%%%%%%%%%%%%%%%%%%%%%%%%%%%%%%
\begin{acks}
 We would like to thank the referee for his/her constructive comments and suggestions which improved the manuscript considerably.
SDO is a mission for NASA's Living With a Star (LWS) Program.
We are thankful for the radio data obtained from NRH.
PK is highly thankful to Prof. P.K. Manoharan for the helpful discussions on radio observations. This work is supported by the ``Development of Korea Space Weather Center" project, and the KASI basic research fund.
\end{acks}
%%%%%%%%%%%%%%%%%%%%%%%%%%%%%%%%%%%%%%%%%%%%%%%%%%%%%%%%%%%%%%%%%%%%%%%%%%%%%
\bibliographystyle{spr-mp-sola}
\bibliography{sun2}

\begin{thebibliography}{34}
% BibTex style file: spr-mp-sola.bst, 2009-06-12
\ifx \bisbn   \undefined \def \bisbn  #1{ISBN #1}\fi
\ifx \binits  \undefined \def \binits#1{#1}\fi
\ifx \bauthor  \undefined \def \bauthor#1{#1}\fi
\ifx \batitle  \undefined \def \batitle#1{#1}\fi
\ifx \bjtitle  \undefined \def \bjtitle#1{\textit{#1}}\fi
\ifx \bvolume  \undefined \def \bvolume#1{\textbf{#1}}\fi
\ifx \byear  \undefined \def \byear#1{#1}\fi
\ifx \bissue  \undefined \def \bissue#1{#1}\fi
\ifx \bfpage  \undefined \def \bfpage#1{#1}\fi
\ifx \blpage  \undefined \def \blpage #1{#1}\fi
\ifx \burl  \undefined \def \burl#1{\textsf{#1}}\fi
\ifx \href  \undefined \def \href#1#2{\textsf{#2}}\fi
\ifx \doiurl  \undefined \def
  \doiurl#1{\href{http://dx.doi.org/#1}{\textsf{#1}}}\fi
\ifx \betal  \undefined \def \betal{\textit{et al.}}\fi
\ifx \binstitute  \undefined \def \binstitute#1{#1}\fi
\ifx \bctitle  \undefined \def \bctitle#1{#1}\fi
\ifx \beditor  \undefined \def \beditor#1{#1}\fi
\ifx \bpublisher  \undefined \def \bpublisher#1{#1}\fi
\ifx \bbtitle  \undefined \def \bbtitle#1{\textit{#1}}\fi
\ifx \bedition  \undefined \def \bedition#1{#1}\fi
\ifx \bseriesno  \undefined \def \bseriesno#1{\textbf{#1}}\fi
\ifx \blocation  \undefined \def \blocation#1{#1}\fi
\ifx \bsertitle  \undefined \def \bsertitle#1{\textit{#1}}\fi
\ifx \bsnm \undefined \def \bsnm#1{#1}\fi
\ifx \bsuffix \undefined \def \bsuffix#1{#1}\fi
\ifx \bparticle \undefined \def \bparticle#1{#1}\fi
\ifx \barticle \undefined \def \barticle{}\fi
\ifx \botherref \undefined \def \botherref{}\fi
\ifx \url \undefined \def \url#1{\textsf{#1}}\fi
\ifx \bchapter \undefined \def \bchapter{}\fi
\ifx \bbook \undefined \def \bbook{}\fi
\ifx \bcomment \undefined \def \bcomment#1{#1}\fi
\ifx \oauthor \undefined \def \oauthor#1{#1}\fi
\ifx \citeauthoryear \undefined \def \citeauthoryear#1{#1}\fi
\def \endbibitem {}

\bibitem[\protect\citeauthoryear{{Antiochos}, {DeVore}, and
  {Klimchuk}}{1999}]{Antiochos99}
\begin{barticle}
\bauthor{\bsnm{{Antiochos}}, \binits{S.K.}}, \bauthor{\bsnm{{DeVore}},
  \binits{C.R.}}, \bauthor{\bsnm{{Klimchuk}}, \binits{J.A.}}:
\byear{1999},
\batitle{{A model for solar coronal mass ejections}}.
\bjtitle{\apj}
\bvolume{510},
\bfpage{485}\,--\,\blpage{493}.
doi:\doiurl{10.1086/306563}.
\end{barticle}
\endbibitem

\bibitem[\protect\citeauthoryear{{Bain} \textit{et~al.}}{2012}]{Bain12}
\begin{barticle}
\bauthor{\bsnm{{Bain}}, \binits{H.M.}}, \bauthor{\bsnm{{Krucker}},
  \binits{S.}}, \bauthor{\bsnm{{Glesener}}, \binits{L.}},
  \bauthor{\bsnm{{Lin}}, \binits{R.P.}}:
\byear{2012},
\batitle{{Radio imaging of shock-accelerated electrons associated with an
  erupting plasmoid on 2010 November 3}}.
\bjtitle{\apj}
\bvolume{750},
\bfpage{44}.
doi:\doiurl{10.1088/0004-637X/750/1/44}.
\end{barticle}
\endbibitem

\bibitem[\protect\citeauthoryear{{Carmichael}}{1964}]{Car64}
\begin{botherref}
\oauthor{\bsnm{{Carmichael}}, \binits{H.}}:
1964,
{A process for flares}.
In: \textit{Hess, W.N. (ed.), The Physics of Solar Flares,}
\textbf{50},
451\,--\,456.
\end{botherref}
\endbibitem

\bibitem[\protect\citeauthoryear{{Cheng} \textit{et~al.}}{2011}]{Cheng11}
\begin{barticle}
\bauthor{\bsnm{{Cheng}}, \binits{X.}}, \bauthor{\bsnm{{Zhang}}, \binits{J.}},
  \bauthor{\bsnm{{Liu}}, \binits{Y.}}, \bauthor{\bsnm{{Ding}}, \binits{M.D.}}:
\byear{2011},
\batitle{{Observing flux rope formation during the impulsive phase of a solar
  eruption}}.
\bjtitle{Astrophys. J. Lett.}
\bvolume{732},
\bfpage{L25}.
doi:\doiurl{10.1088/2041-8205/732/2/L25}.
\end{barticle}
\endbibitem

\bibitem[\protect\citeauthoryear{{Foullon} \textit{et~al.}}{2011}]{Foullon11}
\begin{barticle}
\bauthor{\bsnm{{Foullon}}, \binits{C.}}, \bauthor{\bsnm{{Verwichte}},
  \binits{E.}}, \bauthor{\bsnm{{Nakariakov}}, \binits{V.M.}},
  \bauthor{\bsnm{{Nykyri}}, \binits{K.}}, \bauthor{\bsnm{{Farrugia}},
  \binits{C.J.}}:
\byear{2011},
\batitle{{Magnetic Kelvin-Helmholtz instability at the Sun}}.
\bjtitle{Astrophys. J. Lett.}
\bvolume{729},
\bfpage{L8}.
doi:\doiurl{10.1088/2041-8205/729/1/L8}.
\end{barticle}
\endbibitem

\bibitem[\protect\citeauthoryear{{Gopalswamy}
  \textit{et~al.}}{2012}]{gopalswamy2012}
\begin{barticle}
\bauthor{\bsnm{{Gopalswamy}}, \binits{N.}}, \bauthor{\bsnm{{Nitta}},
  \binits{N.}}, \bauthor{\bsnm{{Akiyama}}, \binits{S.}},
  \bauthor{\bsnm{{M{\"a}kel{\"a}}}, \binits{P.}}, \bauthor{\bsnm{{Yashiro}},
  \binits{S.}}:
\byear{2012},
\batitle{{Coronal magnetic field measurement from EUV images made by the Solar
  Dynamics Observatory}}.
\bjtitle{\apj}
\bvolume{744},
\bfpage{72}.
doi:\doiurl{10.1088/0004-637X/744/1/72}.
\end{barticle}
\endbibitem

\bibitem[\protect\citeauthoryear{{Hannah} and {Kontar}}{2012}]{Hannah12}
\begin{botherref}
\oauthor{\bsnm{{Hannah}}, \binits{I.G.}}, \oauthor{\bsnm{{Kontar}},
  \binits{E.P.}}:
2012,
{Multi-thermal dynamics and energetics of a coronal mass ejection in the low
  solar atmosphere}.
\textit{ArXiv e-prints}.
\end{botherref}
\endbibitem

\bibitem[\protect\citeauthoryear{{Hirayama}}{1985}]{Hira85}
\begin{barticle}
\bauthor{\bsnm{{Hirayama}}, \binits{T.}}:
\byear{1985},
\batitle{{Modern observations of solar prominences}}.
\bjtitle{\solphys}
\bvolume{100},
\bfpage{415}\,--\,\blpage{434}.
doi:\doiurl{10.1007/BF00158439}.
\end{barticle}
\endbibitem

\bibitem[\protect\citeauthoryear{{Howard} \textit{et~al.}}{2008}]{Howard08}
\begin{barticle}
\bauthor{\bsnm{{Howard}}, \binits{R.A.}}, \bauthor{\bsnm{{Moses}},
  \binits{J.D.}}, \bauthor{\bsnm{{Vourlidas}}, \binits{A.}},
  \bauthor{\bsnm{{Newmark}}, \binits{J.S.}}, \bauthor{\bsnm{{Socker}},
  \binits{D.G.}}, \bauthor{\bsnm{{Plunkett}}, \binits{S.P.}},
  \bauthor{\bsnm{{Korendyke}}, \binits{C.M.}}, \bauthor{\bsnm{{Cook}},
  \binits{J.W.}}, \bauthor{\bsnm{{Hurley}}, \binits{A.}},
  \bauthor{\bsnm{{Davila}}, \binits{J.M.}}, \bauthor{\bsnm{{Thompson}},
  \binits{W.T.}}, \bauthor{\bsnm{{St Cyr}}, \binits{O.C.}},
  \bauthor{\bsnm{{Mentzell}}, \binits{E.}}, \bauthor{\bsnm{{Mehalick}},
  \binits{K.}}, \bauthor{\bsnm{{Lemen}}, \binits{J.R.}},
  \bauthor{\bsnm{{Wuelser}}, \binits{J.P.}}, \bauthor{\bsnm{{Duncan}},
  \binits{D.W.}}, \bauthor{\bsnm{{Tarbell}}, \binits{T.D.}},
  \bauthor{\bsnm{{Wolfson}}, \binits{C.J.}}, \bauthor{\bsnm{{Moore}},
  \binits{A.}}, \bauthor{\bsnm{{Harrison}}, \binits{R.A.}},
  \bauthor{\bsnm{{Waltham}}, \binits{N.R.}}, \bauthor{\bsnm{{Lang}},
  \binits{J.}}, \bauthor{\bsnm{{Davis}}, \binits{C.J.}},
  \bauthor{\bsnm{{Eyles}}, \binits{C.J.}}, \bauthor{\bsnm{{Mapson-Menard}},
  \binits{H.}}, \bauthor{\bsnm{{Simnett}}, \binits{G.M.}},
  \bauthor{\bsnm{{Halain}}, \binits{J.P.}}, \bauthor{\bsnm{{Defise}},
  \binits{J.M.}}, \bauthor{\bsnm{{Mazy}}, \binits{E.}},
  \bauthor{\bsnm{{Rochus}}, \binits{P.}}, \bauthor{\bsnm{{Mercier}},
  \binits{R.}}, \bauthor{\bsnm{{Ravet}}, \binits{M.F.}},
  \bauthor{\bsnm{{Delmotte}}, \binits{F.}}, \bauthor{\bsnm{{Auchere}},
  \binits{F.}}, \bauthor{\bsnm{{Delaboudiniere}}, \binits{J.P.}},
  \bauthor{\bsnm{{Bothmer}}, \binits{V.}}, \bauthor{\bsnm{{Deutsch}},
  \binits{W.}}, \bauthor{\bsnm{{Wang}}, \binits{D.}}, \bauthor{\bsnm{{Rich}},
  \binits{N.}}, \bauthor{\bsnm{{Cooper}}, \binits{S.}},
  \bauthor{\bsnm{{Stephens}}, \binits{V.}}, \bauthor{\bsnm{{Maahs}},
  \binits{G.}}, \bauthor{\bsnm{{Baugh}}, \binits{R.}},
  \bauthor{\bsnm{{McMullin}}, \binits{D.}}, \bauthor{\bsnm{{Carter}},
  \binits{T.}}:
\byear{2008},
\batitle{{Sun Earth Connection Coronal and Heliospheric Investigation
  (SECCHI)}}.
\bjtitle{Space Sci. Rev.}
\bvolume{136},
\bfpage{67}\,--\,\blpage{115}.
doi:\doiurl{10.1007/s11214-008-9341-4}.
\end{barticle}
\endbibitem

\bibitem[\protect\citeauthoryear{{Innes} \textit{et~al.}}{2001}]{Ietal01}
\begin{barticle}
\bauthor{\bsnm{{Innes}}, \binits{D.E.}}, \bauthor{\bsnm{{Curdt}}, \binits{W.}},
  \bauthor{\bsnm{{Schwenn}}, \binits{R.}}, \bauthor{\bsnm{{Solanki}},
  \binits{S.}}, \bauthor{\bsnm{{Stenborg}}, \binits{G.}},
  \bauthor{\bsnm{{McKenzie}}, \binits{D.E.}}:
\byear{2001},
\batitle{{Large Doppler shifts in X-ray plasma: An explosive start to coronal
  mass ejection}}.
\bjtitle{Astrophys. J. Lett.}
\bvolume{549},
\bfpage{L249}\,--\,\blpage{L252}.
doi:\doiurl{10.1086/319164}.
\end{barticle}
\endbibitem

\bibitem[\protect\citeauthoryear{{Karpen}, {Antiochos}, and
  {DeVore}}{2012}]{Karpen12}
\begin{barticle}
\bauthor{\bsnm{{Karpen}}, \binits{J.T.}}, \bauthor{\bsnm{{Antiochos}},
  \binits{S.K.}}, \bauthor{\bsnm{{DeVore}}, \binits{C.R.}}:
\byear{2012},
\batitle{{The mechanisms for the onset and explosive eruption of coronal mass
  ejections and eruptive flares}}.
\bjtitle{\apj}
\bvolume{760},
\bfpage{81}.
doi:\doiurl{10.1088/0004-637X/760/1/81}.
\end{barticle}
\endbibitem

\bibitem[\protect\citeauthoryear{{Kerdraon} and {Delouis}}{1997}]{Kerdraon1997}
\begin{botherref}
\oauthor{\bsnm{{Kerdraon}}, \binits{A.}}, \oauthor{\bsnm{{Delouis}},
  \binits{J.M.}}:
1997,
{The Nan{\c c}ay Radioheliograph}.
In: {Trottet}, G. (ed.)
\textit{Coronal Physics from Radio and Space Observations},
\textit{Lecture Notes in Physics, Springer Verlag, Berlin,}
\textbf{483},
192\,--\,201.
doi:\doiurl{10.1007/BFb0106458}.
\end{botherref}
\endbibitem

\bibitem[\protect\citeauthoryear{{Kopp} and {Pneuman}}{1976}]{KP76}
\begin{barticle}
\bauthor{\bsnm{{Kopp}}, \binits{R.A.}}, \bauthor{\bsnm{{Pneuman}},
  \binits{G.W.}}:
\byear{1976},
\batitle{{Magnetic reconnection in the corona and the loop prominence
  phenomenon}}.
\bjtitle{\solphys}
\bvolume{50},
\bfpage{85}\,--\,\blpage{98}.
doi:\doiurl{10.1007/BF00206193}.
\end{barticle}
\endbibitem

\bibitem[\protect\citeauthoryear{{Kozarev} \textit{et~al.}}{2011}]{kozarev2011}
\begin{barticle}
\bauthor{\bsnm{{Kozarev}}, \binits{K.A.}}, \bauthor{\bsnm{{Korreck}},
  \binits{K.E.}}, \bauthor{\bsnm{{Lobzin}}, \binits{V.V.}},
  \bauthor{\bsnm{{Weber}}, \binits{M.A.}}, \bauthor{\bsnm{{Schwadron}},
  \binits{N.A.}}:
\byear{2011},
\batitle{{Off-limb solar coronal wavefronts from SDO/AIA Extreme-ultraviolet
  Observations: Implications for particle production}}.
\bjtitle{Astrophys. J. Lett.}
\bvolume{733},
\bfpage{L25}.
doi:\doiurl{10.1088/2041-8205/733/2/L25}.
\end{barticle}
\endbibitem

\bibitem[\protect\citeauthoryear{{Lemen} \textit{et~al.}}{2012}]{Lemen12}
\begin{barticle}
\bauthor{\bsnm{{Lemen}}, \binits{J.R.}}, \bauthor{\bsnm{{Title}},
  \binits{A.M.}}, \bauthor{\bsnm{{Akin}}, \binits{D.J.}},
  \bauthor{\bsnm{{Boerner}}, \binits{P.F.}}, \bauthor{\bsnm{{Chou}},
  \binits{C.}}, \bauthor{\bsnm{{Drake}}, \binits{J.F.}},
  \bauthor{\bsnm{{Duncan}}, \binits{D.W.}}, \bauthor{\bsnm{{Edwards}},
  \binits{C.G.}}, \bauthor{\bsnm{{Friedlaender}}, \binits{F.M.}},
  \bauthor{\bsnm{{Heyman}}, \binits{G.F.}}, \bauthor{\bsnm{{Hurlburt}},
  \binits{N.E.}}, \bauthor{\bsnm{{Katz}}, \binits{N.L.}},
  \bauthor{\bsnm{{Kushner}}, \binits{G.D.}}, \bauthor{\bsnm{{Levay}},
  \binits{M.}}, \bauthor{\bsnm{{Lindgren}}, \binits{R.W.}},
  \bauthor{\bsnm{{Mathur}}, \binits{D.P.}}, \bauthor{\bsnm{{McFeaters}},
  \binits{E.L.}}, \bauthor{\bsnm{{Mitchell}}, \binits{S.}},
  \bauthor{\bsnm{{Rehse}}, \binits{R.A.}}, \bauthor{\bsnm{{Schrijver}},
  \binits{C.J.}}, \bauthor{\bsnm{{Springer}}, \binits{L.A.}},
  \bauthor{\bsnm{{Stern}}, \binits{R.A.}}, \bauthor{\bsnm{{Tarbell}},
  \binits{T.D.}}, \bauthor{\bsnm{{Wuelser}}, \binits{J.P.}},
  \bauthor{\bsnm{{Wolfson}}, \binits{C.J.}}, \bauthor{\bsnm{{Yanari}},
  \binits{C.}}, \bauthor{\bsnm{{Bookbinder}}, \binits{J.A.}},
  \bauthor{\bsnm{{Cheimets}}, \binits{P.N.}}, \bauthor{\bsnm{{Caldwell}},
  \binits{D.}}, \bauthor{\bsnm{{Deluca}}, \binits{E.E.}},
  \bauthor{\bsnm{{Gates}}, \binits{R.}}, \bauthor{\bsnm{{Golub}}, \binits{L.}},
  \bauthor{\bsnm{{Park}}, \binits{S.}}, \bauthor{\bsnm{{Podgorski}},
  \binits{W.A.}}, \bauthor{\bsnm{{Bush}}, \binits{R.I.}},
  \bauthor{\bsnm{{Scherrer}}, \binits{P.H.}}, \bauthor{\bsnm{{Gummin}},
  \binits{M.A.}}, \bauthor{\bsnm{{Smith}}, \binits{P.}},
  \bauthor{\bsnm{{Auker}}, \binits{G.}}, \bauthor{\bsnm{{Jerram}},
  \binits{P.}}, \bauthor{\bsnm{{Pool}}, \binits{P.}}, \bauthor{\bsnm{{Soufli}},
  \binits{R.}}, \bauthor{\bsnm{{Windt}}, \binits{D.L.}},
  \bauthor{\bsnm{{Beardsley}}, \binits{S.}}, \bauthor{\bsnm{{Clapp}},
  \binits{M.}}, \bauthor{\bsnm{{Lang}}, \binits{J.}},
  \bauthor{\bsnm{{Waltham}}, \binits{N.}}:
\byear{2012},
\batitle{{The Atmospheric Imaging Assembly (AIA) on the Solar Dynamics
  Observatory (SDO)}}.
\bjtitle{\solphys}
\bvolume{275},
\bfpage{17}\,--\,\blpage{40}.
doi:\doiurl{10.1007/s11207-011-9776-8}.
\end{barticle}
\endbibitem

\bibitem[\protect\citeauthoryear{{Lin} \textit{et~al.}}{2002}]{lin2002}
\begin{barticle}
\bauthor{\bsnm{{Lin}}, \binits{R.P.}}, \bauthor{\bsnm{{Dennis}},
  \binits{B.R.}}, \bauthor{\bsnm{{Hurford}}, \binits{G.J.}},
  \bauthor{\bsnm{{Smith}}, \binits{D.M.}}, \bauthor{\bsnm{{Zehnder}},
  \binits{A.}}, \bauthor{\bsnm{{Harvey}}, \binits{P.R.}},
  \bauthor{\bsnm{{Curtis}}, \binits{D.W.}}, \bauthor{\bsnm{{Pankow}},
  \binits{D.}}, \bauthor{\bsnm{{Turin}}, \binits{P.}},
  \bauthor{\bsnm{{Bester}}, \binits{M.}}, \bauthor{\bsnm{{Csillaghy}},
  \binits{A.}}, \bauthor{\bsnm{{Lewis}}, \binits{M.}},
  \bauthor{\bsnm{{Madden}}, \binits{N.}}, \bauthor{\bsnm{{van Beek}},
  \binits{H.F.}}, \bauthor{\bsnm{{Appleby}}, \binits{M.}},
  \bauthor{\bsnm{{Raudorf}}, \binits{T.}}, \bauthor{\bsnm{{McTiernan}},
  \binits{J.}}, \bauthor{\bsnm{{Ramaty}}, \binits{R.}},
  \bauthor{\bsnm{{Schmahl}}, \binits{E.}}, \bauthor{\bsnm{{Schwartz}},
  \binits{R.}}, \bauthor{\bsnm{{Krucker}}, \binits{S.}},
  \bauthor{\bsnm{{Abiad}}, \binits{R.}}, \bauthor{\bsnm{{Quinn}}, \binits{T.}},
  \bauthor{\bsnm{{Berg}}, \binits{P.}}, \bauthor{\bsnm{{Hashii}}, \binits{M.}},
  \bauthor{\bsnm{{Sterling}}, \binits{R.}}, \bauthor{\bsnm{{Jackson}},
  \binits{R.}}, \bauthor{\bsnm{{Pratt}}, \binits{R.}},
  \bauthor{\bsnm{{Campbell}}, \binits{R.D.}}, \bauthor{\bsnm{{Malone}},
  \binits{D.}}, \bauthor{\bsnm{{Landis}}, \binits{D.}},
  \bauthor{\bsnm{{Barrington-Leigh}}, \binits{C.P.}},
  \bauthor{\bsnm{{Slassi-Sennou}}, \binits{S.}}, \bauthor{\bsnm{{Cork}},
  \binits{C.}}, \bauthor{\bsnm{{Clark}}, \binits{D.}}, \bauthor{\bsnm{{Amato}},
  \binits{D.}}, \bauthor{\bsnm{{Orwig}}, \binits{L.}}, \bauthor{\bsnm{{Boyle}},
  \binits{R.}}, \bauthor{\bsnm{{Banks}}, \binits{I.S.}},
  \bauthor{\bsnm{{Shirey}}, \binits{K.}}, \bauthor{\bsnm{{Tolbert}},
  \binits{A.K.}}, \bauthor{\bsnm{{Zarro}}, \binits{D.}},
  \bauthor{\bsnm{{Snow}}, \binits{F.}}, \bauthor{\bsnm{{Thomsen}},
  \binits{K.}}, \bauthor{\bsnm{{Henneck}}, \binits{R.}},
  \bauthor{\bsnm{{McHedlishvili}}, \binits{A.}}, \bauthor{\bsnm{{Ming}},
  \binits{P.}}, \bauthor{\bsnm{{Fivian}}, \binits{M.}},
  \bauthor{\bsnm{{Jordan}}, \binits{J.}}, \bauthor{\bsnm{{Wanner}},
  \binits{R.}}, \bauthor{\bsnm{{Crubb}}, \binits{J.}},
  \bauthor{\bsnm{{Preble}}, \binits{J.}}, \bauthor{\bsnm{{Matranga}},
  \binits{M.}}, \bauthor{\bsnm{{Benz}}, \binits{A.}}, \bauthor{\bsnm{{Hudson}},
  \binits{H.}}, \bauthor{\bsnm{{Canfield}}, \binits{R.C.}},
  \bauthor{\bsnm{{Holman}}, \binits{G.D.}}, \bauthor{\bsnm{{Crannell}},
  \binits{C.}}, \bauthor{\bsnm{{Kosugi}}, \binits{T.}},
  \bauthor{\bsnm{{Emslie}}, \binits{A.G.}}, \bauthor{\bsnm{{Vilmer}},
  \binits{N.}}, \bauthor{\bsnm{{Brown}}, \binits{J.C.}},
  \bauthor{\bsnm{{Johns-Krull}}, \binits{C.}}, \bauthor{\bsnm{{Aschwanden}},
  \binits{M.}}, \bauthor{\bsnm{{Metcalf}}, \binits{T.}},
  \bauthor{\bsnm{{Conway}}, \binits{A.}}:
\byear{2002},
\batitle{{The Reuven Ramaty High-Energy Solar Spectroscopic Imager (RHESSI)}}.
\bjtitle{\solphys}
\bvolume{210},
\bfpage{3}\,--\,\blpage{32}.
doi:\doiurl{10.1023/A:1022428818870}.
\end{barticle}
\endbibitem

\bibitem[\protect\citeauthoryear{{Ma} \textit{et~al.}}{2011}]{ma2011}
\begin{barticle}
\bauthor{\bsnm{{Ma}}, \binits{S.}}, \bauthor{\bsnm{{Raymond}}, \binits{J.C.}},
  \bauthor{\bsnm{{Golub}}, \binits{L.}}, \bauthor{\bsnm{{Lin}}, \binits{J.}},
  \bauthor{\bsnm{{Chen}}, \binits{H.}}, \bauthor{\bsnm{{Grigis}}, \binits{P.}},
  \bauthor{\bsnm{{Testa}}, \binits{P.}}, \bauthor{\bsnm{{Long}}, \binits{D.}}:
\byear{2011},
\batitle{{Observations and interpretation of a low coronal shock wave observed
  in the EUV by the SDO/AIA}}.
\bjtitle{\apj}
\bvolume{738},
\bfpage{160}.
doi:\doiurl{10.1088/0004-637X/738/2/160}.
\end{barticle}
\endbibitem

\bibitem[\protect\citeauthoryear{{Magdaleni{\'c}}
  \textit{et~al.}}{2008}]{magdalenic2008}
\begin{barticle}
\bauthor{\bsnm{{Magdaleni{\'c}}}, \binits{J.}}, \bauthor{\bsnm{{Vr{\v s}nak}},
  \binits{B.}}, \bauthor{\bsnm{{Pohjolainen}}, \binits{S.}},
  \bauthor{\bsnm{{Temmer}}, \binits{M.}}, \bauthor{\bsnm{{Aurass}},
  \binits{H.}}, \bauthor{\bsnm{{Lehtinen}}, \binits{N.J.}}:
\byear{2008},
\batitle{{A flare-generated shock during a coronal mass ejection on 24 December
  1996}}.
\bjtitle{\solphys}
\bvolume{253},
\bfpage{305}\,--\,\blpage{317}.
doi:\doiurl{10.1007/s11207-008-9220-x}.
\end{barticle}
\endbibitem

\bibitem[\protect\citeauthoryear{{Magdaleni{\'c}}
  \textit{et~al.}}{2010}]{magdalenic2010}
\begin{barticle}
\bauthor{\bsnm{{Magdaleni{\'c}}}, \binits{J.}}, \bauthor{\bsnm{{Marqu{\'e}}},
  \binits{C.}}, \bauthor{\bsnm{{Zhukov}}, \binits{A.N.}}, \bauthor{\bsnm{{Vr{\v
  s}nak}}, \binits{B.}}, \bauthor{\bsnm{{{\v Z}ic}}, \binits{T.}}:
\byear{2010},
\batitle{{Origin of coronal shock waves associated with slow coronal mass
  ejections}}.
\bjtitle{\apj}
\bvolume{718},
\bfpage{266}\,--\,\blpage{278}.
doi:\doiurl{10.1088/0004-637X/718/1/266}.
\end{barticle}
\endbibitem

\bibitem[\protect\citeauthoryear{{Magdaleni{\'c}}
  \textit{et~al.}}{2012}]{Magdalenic12}
\begin{barticle}
\bauthor{\bsnm{{Magdaleni{\'c}}}, \binits{J.}}, \bauthor{\bsnm{{Marqu{\'e}}},
  \binits{C.}}, \bauthor{\bsnm{{Zhukov}}, \binits{A.N.}}, \bauthor{\bsnm{{Vr{\v
  s}nak}}, \binits{B.}}, \bauthor{\bsnm{{Veronig}}, \binits{A.}}:
\byear{2012},
\batitle{{Flare-generated type II burst without associated coronal mass
  ejection}}.
\bjtitle{\apj}
\bvolume{746},
\bfpage{152}.
doi:\doiurl{10.1088/0004-637X/746/2/152}.
\end{barticle}
\endbibitem

\bibitem[\protect\citeauthoryear{{Patsourakos} and
  {Vourlidas}}{2012}]{patsourakos2012}
\begin{barticle}
\bauthor{\bsnm{{Patsourakos}}, \binits{S.}}, \bauthor{\bsnm{{Vourlidas}},
  \binits{A.}}:
\byear{2012},
\batitle{{On the nature and genesis of EUV waves: A synthesis of observations
  from SOHO, STEREO, SDO, and Hinode (Invited review)}}.
\bjtitle{\solphys}
\bvolume{281},
\bfpage{187}\,--\,\blpage{222}.
doi:\doiurl{10.1007/s11207-012-9988-6}.
\end{barticle}
\endbibitem

\bibitem[\protect\citeauthoryear{{Pesnell}, {Thompson}, and
  {Chamberlin}}{2012}]{Pesnell2012}
\begin{barticle}
\bauthor{\bsnm{{Pesnell}}, \binits{W.D.}}, \bauthor{\bsnm{{Thompson}},
  \binits{B.J.}}, \bauthor{\bsnm{{Chamberlin}}, \binits{P.C.}}:
\byear{2012},
\batitle{{The Solar Dynamics Observatory (SDO)}}.
\bjtitle{\solphys}
\bvolume{275},
\bfpage{3}\,--\,\blpage{15}.
doi:\doiurl{10.1007/s11207-011-9841-3}.
\end{barticle}
\endbibitem

\bibitem[\protect\citeauthoryear{{Reeves} and {Golub}}{2011}]{Reeves11}
\begin{barticle}
\bauthor{\bsnm{{Reeves}}, \binits{K.K.}}, \bauthor{\bsnm{{Golub}},
  \binits{L.}}:
\byear{2011},
\batitle{{Atmospheric Imaging Assembly observations of hot flare plasma}}.
\bjtitle{Astrophys. J. Lett.}
\bvolume{727},
\bfpage{L52}.
doi:\doiurl{10.1088/2041-8205/727/2/L52}.
\end{barticle}
\endbibitem

\bibitem[\protect\citeauthoryear{{Savage} \textit{et~al.}}{2012}]{Savage12}
\begin{barticle}
\bauthor{\bsnm{{Savage}}, \binits{S.L.}}, \bauthor{\bsnm{{Holman}},
  \binits{G.}}, \bauthor{\bsnm{{Reeves}}, \binits{K.K.}},
  \bauthor{\bsnm{{Seaton}}, \binits{D.B.}}, \bauthor{\bsnm{{McKenzie}},
  \binits{D.E.}}, \bauthor{\bsnm{{Su}}, \binits{Y.}}:
\byear{2012},
\batitle{{Low-altitude reconnection inflow-outflow observations during a 2010
  November 3 solar eruption}}.
\bjtitle{\apj}
\bvolume{754},
\bfpage{13}.
doi:\doiurl{10.1088/0004-637X/754/1/13}.
\end{barticle}
\endbibitem

\bibitem[\protect\citeauthoryear{{Sturrock}}{1966}]{Sturrock66}
\begin{barticle}
\bauthor{\bsnm{{Sturrock}}, \binits{P.A.}}:
\byear{1966},
\batitle{{Model of the high-energy phase of solar flares}}.
\bjtitle{\nat}
\bvolume{211},
\bfpage{695}\,--\,\blpage{697}.
doi:\doiurl{10.1038/211695a0}.
\end{barticle}
\endbibitem

\bibitem[\protect\citeauthoryear{{Thompson}}{2006}]{thompson2006}
\begin{barticle}
\bauthor{\bsnm{{Thompson}}, \binits{W.T.}}:
\byear{2006},
\batitle{{Coordinate systems for solar image data}}.
\bjtitle{\aap}
\bvolume{449},
\bfpage{791}\,--\,\blpage{803}.
doi:\doiurl{10.1051/0004-6361:20054262}.
\end{barticle}
\endbibitem

\bibitem[\protect\citeauthoryear{{Tothova}, {Innes}, and
  {Stenborg}}{2011}]{Tothova11}
\begin{barticle}
\bauthor{\bsnm{{Tothova}}, \binits{D.}}, \bauthor{\bsnm{{Innes}},
  \binits{D.E.}}, \bauthor{\bsnm{{Stenborg}}, \binits{G.}}:
\byear{2011},
\batitle{{Oscillations in the wake of a flare blast wave}}.
\bjtitle{\aap}
\bvolume{528},
\bfpage{L12}.
doi:\doiurl{10.1051/0004-6361/201015272}.
\end{barticle}
\endbibitem

\bibitem[\protect\citeauthoryear{{Vr{\v s}nak} and {Cliver}}{2008}]{vrsnak2008}
\begin{barticle}
\bauthor{\bsnm{{Vr{\v s}nak}}, \binits{B.}}, \bauthor{\bsnm{{Cliver}},
  \binits{E.W.}}:
\byear{2008},
\batitle{{Origin of coronal shock waves. Invited review}}.
\bjtitle{\solphys}
\bvolume{253},
\bfpage{215}\,--\,\blpage{235}.
doi:\doiurl{10.1007/s11207-008-9241-5}.
\end{barticle}
\endbibitem

\bibitem[\protect\citeauthoryear{{Vr{\v s}nak} and
  {Luli{\'c}}}{2000a}]{Vrsnak00b}
\begin{barticle}
\bauthor{\bsnm{{Vr{\v s}nak}}, \binits{B.}}, \bauthor{\bsnm{{Luli{\'c}}},
  \binits{S.}}:
\byear{2000}a,
\batitle{{Formation of coronal MHD shock waves - I. The basic mechanism}}.
\bjtitle{\solphys}
\bvolume{196},
\bfpage{157}\,--\,\blpage{180}.
doi:\doiurl{10.1023/A:100523680472}.
\end{barticle}
\endbibitem

\bibitem[\protect\citeauthoryear{{Vr{\v s}nak} and
  {Luli{\'c}}}{2000b}]{Vrsnak00a}
\begin{barticle}
\bauthor{\bsnm{{Vr{\v s}nak}}, \binits{B.}}, \bauthor{\bsnm{{Luli{\'c}}},
  \binits{S.}}:
\byear{2000}b,
\batitle{{Formation of coronal MHD shock waves - II. The pressure pulse
  mechanism}}.
\bjtitle{\solphys}
\bvolume{196},
\bfpage{181}\,--\,\blpage{197}.
doi:\doiurl{10.1023/A:1005288310697}.
\end{barticle}
\endbibitem

\bibitem[\protect\citeauthoryear{{Vr{\v s}nak}
  \textit{et~al.}}{2006}]{Vrsnal06}
\begin{barticle}
\bauthor{\bsnm{{Vr{\v s}nak}}, \binits{B.}}, \bauthor{\bsnm{{Warmuth}},
  \binits{A.}}, \bauthor{\bsnm{{Temmer}}, \binits{M.}},
  \bauthor{\bsnm{{Veronig}}, \binits{A.}}, \bauthor{\bsnm{{Magdaleni{\'c}}},
  \binits{J.}}, \bauthor{\bsnm{{Hillaris}}, \binits{A.}},
  \bauthor{\bsnm{{Karlick{\'y}}}, \binits{M.}}:
\byear{2006},
\batitle{{Multi-wavelength study of coronal waves associated with the CME-flare
  event of 3 November 2003}}.
\bjtitle{\aap}
\bvolume{448},
\bfpage{739}\,--\,\blpage{752}.
doi:\doiurl{10.1051/0004-6361:20053740}.
\end{barticle}
\endbibitem

\bibitem[\protect\citeauthoryear{{White} and {Verwichte}}{2012}]{White12a}
\begin{barticle}
\bauthor{\bsnm{{White}}, \binits{R.S.}}, \bauthor{\bsnm{{Verwichte}},
  \binits{E.}}:
\byear{2012},
\batitle{{Transverse coronal loop oscillations seen in unprecedented detail by
  AIA/SDO}}.
\bjtitle{\aap}
\bvolume{537},
\bfpage{A49}.
doi:\doiurl{10.1051/0004-6361/201118093}.
\end{barticle}
\endbibitem

\bibitem[\protect\citeauthoryear{{White}, {Verwichte}, and
  {Foullon}}{2012}]{White12b}
\begin{barticle}
\bauthor{\bsnm{{White}}, \binits{R.S.}}, \bauthor{\bsnm{{Verwichte}},
  \binits{E.}}, \bauthor{\bsnm{{Foullon}}, \binits{C.}}:
\byear{2012},
\batitle{{First observation of a transverse vertical oscillation during the
  formation of a hot post-flare loop}}.
\bjtitle{\aap}
\bvolume{545},
\bfpage{A129}.
doi:\doiurl{10.1051/0004-6361/201219856}.
\end{barticle}
\endbibitem

\bibitem[\protect\citeauthoryear{{Zimovets} \textit{et~al.}}{2012}]{Zimovets12}
\begin{barticle}
\bauthor{\bsnm{{Zimovets}}, \binits{I.}}, \bauthor{\bsnm{{Vilmer}},
  \binits{N.}}, \bauthor{\bsnm{{Chian}}, \binits{A.C.L.}},
  \bauthor{\bsnm{{Sharykin}}, \binits{I.}}, \bauthor{\bsnm{{Struminsky}},
  \binits{A.}}:
\byear{2012},
\batitle{{Spatially resolved observations of a split-band coronal type II radio
  burst}}.
\bjtitle{\aap}
\bvolume{547},
\bfpage{A6}.
doi:\doiurl{10.1051/0004-6361/201219454}.
\end{barticle}
\endbibitem

\end{thebibliography}
%\bibliography{reference}
\end{article}
\end{document}